\documentclass[%
reprint,
nobibnotes,
 amsmath,amssymb,
longbibliography,
aps,
pre,
floatfix,
]{revtex4-1}

\usepackage{amsmath}
\usepackage{amsthm}
\usepackage{amsfonts}
\usepackage{amssymb}
\usepackage{graphics}
\usepackage{psfrag}
\usepackage{numprint}
\usepackage{notes2bib}
\usepackage{adjustbox}
\usepackage{multirow}
\usepackage{mathtools}
\usepackage{enumitem}
\usepackage{graphicx}
\usepackage{dcolumn}
\usepackage{bm}
\usepackage[colorlinks=true,linkcolor=blue,citecolor=blue,urlcolor=blue]{hyperref}
\usepackage{xcolor}
\usepackage{siunitx}
\usepackage[caption=false]{subfig}
\usepackage[symbol]{footmisc}

\begin{document}
\title{Dynamic Measure of Hyperuniformity and Nonhyperuniformity in Heterogeneous Media via the Diffusion Spreadability}

\author{Haina Wang}
\affiliation{\emph{Department of Chemistry, Princeton University}, Princeton, New Jersey, 08544, USA}
\author{Salvatore Torquato}
\email{torquato@princeton.edu}
\affiliation{\emph{Department of Chemistry, Department of Physics, Princeton Center for Theoretical Science, Princeton Institute for the Science and Technology of Materials, and Program in Applied and Computational Mathematics, Princeton University}, Princeton, New Jersey, 08544 USA}
\date{\today}

\begin{abstract}
Time-dependent interphase diffusion processes in multiphase heterogeneous media arise ubiquitously in physics, chemistry and biology. Examples of heterogeneous media include composites, geological media, gels, foams and cell aggregates. The recently developed concept of spreadability, $\mathcal{S}(t)$, provides a direct link between time-dependent diffusive transport and the microstructure of two-phase media across length scales [Torquato, S., \textit{Phys. Rev. E.}, 104 054102 (2021)]. To investigate the capability of $\mathcal{S}(t)$ to probe microstructures of real heterogeneous media, we explicitly compute $\mathcal{S}(t)$ for well-known two-dimensional and three-dimensional idealized model structures that span across nonhyperuniform and hyperuniform classes. Among the former class, we study fully penetrable spheres and equilibrium hard spheres, and in the latter class, we examine sphere packings derived from ``perfect glasses'', uniformly randomized lattices (URL), disordered stealthy hyperuniform point processes and Bravais lattices. Hyperuniform media are characterized by an anomalous suppression of volume fraction fluctuations at large length scales compared to that of any nonhyperuniform medium. We further confirm that the small-, intermediate- and long-time behaviors of $\mathcal{S}(t)$ sensitively capture the small-, intermediate- and large-scale characteristics of the models. In instances in which the spectral density $\tilde{\chi}_{_V}(\mathbf{k})$ has a power-law form $B|\mathbf{k}|^\alpha$ in the limit $|\mathbf{k}|\rightarrow 0$, the long-time spreadability provides a simple means to extract the value of the coefficients $\alpha$ and $B$ that is robust against noise in $\tilde{\chi}_{_V}(\mathbf{k})$ at small wavenumbers. For typical nonhyperuniform media, the intermediate-time spreadability is slower for models with larger values of the coefficient $B=\tilde{\chi}_{_V}(0)$. Interestingly, the excess spreadability  $\mathcal{S}(\infty)-\mathcal{S}(t)$ for URL packings has nearly exponential decay at small to intermediate $t$, but transforms to a power-law decay at large $t$, and the time for this transition has a logarithmic divergence in the limit of vanishing lattice perturbation. Our study of the aforementioned
models enables us to devise an algorithm that efficiently and accurately extracts large-scale behaviors from diffusion data alone. Lessons learned from such analyses of our models are used to determine accurately the large-scale structural characteristics of a sample Fontainebleau sandstone, which we show is nonhyperuniform. Our study demonstrates the practical applicability of the diffusion spreadability to extract crucial microstructural information from real data across length scales and provides a basis for the inverse design of materials with desirable time-dependent diffusion properties. 
\end{abstract}

\maketitle

\section{Introduction}
\label{intro}

Time-dependent interphase diffusion processes in multiphase heterogeneous media arise in a wide range of physical, chemical and biological contexts, including magnetic resonance imaging \cite{We05}, surface catalysis \cite{To02a, Sosna2020}, material design \cite{To02a,Sa03,Ta18}, cell-behavior modeling \cite{Br79} and controlled drug delivery \cite{La81}. Heterogeneous media are ubiquitous; examples include composites, geological media, gels, foams, cell aggregates, among other natural and synthetic media \cite{To98b,To02a,Mi02,Kl19b,Huang2021}. It is well known that the effective transport properties of heterogeneous media generally depend on an infinite set of correlation functions that characterize the microstructure \cite{To02a,Se89,To21a}.

The {\it spreadability} concept, as very recently developed and explored by Torquato \cite{To21d}, makes a direct link between time-dependent diffusive transport and the microstructure of a given heterogeneous material across length scales. Consider the time-dependent problem of mass transfer of a solute in a two-phase media, where phases 1 and 2 occupy volume fractions $\phi_1$ and $\phi_2$, respectively. Assume that at time $t = 0$, the solute is uniformly distributed throughout phase 2, and completely absent from phase 1. Assume also that the solute has the same diffusion coefficient $D$ in each phase at any $t$. We call the fraction of total solute present in phase 1 as a function of time $\mathcal{S}(t)$ \textit{spreadability}, since it is a measure of the spreadability of diffusion information as a function of time. Generalizing a formula due to Prager in three-dimensional physical
(direct) space \cite{Pr63b}, Torquato showed that $\mathcal{S}(t)$ in any Euclidean space dimension $d$ is exactly related to the microstructure via the autocovariance function $\chi_{_V}(\mathbf{r})$ in direct space, or equivalently, via the spectral density $\tilde{\chi}_{_V}(\mathbf{k})$ in Fourier space  \cite{To21d}:
\begin{widetext}
\begin{subequations}\label{spreadability}
\begin{align}
{\cal S}(\infty)- {\cal S}(t) =& \frac{1}{(4\pi D t)^{d/2}\, \phi_2} \int_{\mathbb{R}^d} \chi_{_V}({\bf r}) \exp[-r^2/(4Dt)] d{\bf r}\label{spreadability_direct} \\
=& \frac{1}{(2\pi)^{d}\phi_2} \int_{\mathbb{R}^d} \tilde{\chi}_{_V}({\bf k}) \exp[-k^2Dt] d{\bf k},\label{spreadability_fourier}
\end{align}
\end{subequations}
\end{widetext}
where $\mathcal{S}(\infty)=\phi_1$ is the infinite-time limit of $\mathcal{S}(t)$, and $\mathcal{S}(\infty)-\mathcal{S}(t)$ is called the \textit{excess spreadability}, i.e., spreadability in excess to its infinite-time value. Equation (\ref{spreadability}) is a singular result because it represents one of the rare examples of transport in two-phase random media where an exact solution is possible only in terms of the first two correlation functions, namely, $\phi_1$ and two-point statistics via either the autocovariance function or spectral density.

Torquato \cite{To21d} has shown that the relation (\ref{spreadability}) implies that small-, intermediate- and long-time behaviors of $\mathcal{S}(t)$ are directly determined by the small-, intermediate- and large-scale structural characteristics of the two-phase medium. Thus, the spreadability has the potential to serve as a simple and powerful dynamic figure of merit to probe and classify all translationally invariant two-phase microstructures that span from hyperuniform to nonhyperuniform media across scales. \textit{Hyperuniform} two-phase media are characterized by an anomalous suppression of volume-fraction fluctuations relative to garden-variety nonhyperuniform disordered media \cite{To03a,Za09,To18a}; see Sec. \ref{hyperuniformity} for exact mathematical definitions. Hyperuniformity is an emerging field, playing vital roles in a number of fundamental and applied contexts, including glass formation \cite{Do05d,Ma13a}, jamming \cite{To07,At16a,Con17,Ri21,To21c}, rigidity \cite{Zhang2016,Gh18}, photonic band-gap materials \cite{Fl09b,Man13a, Fr17}, biology \cite{Ji14,Ma15}, localization of waves and excitations \cite{Fl09a,Ma13b,Zh19,Sgrignuoli2021}, antenna or laser array designs \cite{Ch21}, self-organization \cite{He15,He17b,Ma19}, fluid dynamics \cite{Lei2019b,Ding18,Du21}, quantum systems \cite{Re67,Fe98,To08b,Ab17,Cr19}, random matrices \cite{Dy70,To08b,La19} and pure mathematics \cite{Sa06,Br19a,To18d,To19,Brauchart2020}. Because disordered hyperuniform two-phase media are states of matter that lie between a crystal and a typical liquid, they can be endowed with novel properties \cite{De16,Zh16,To18a,Bi19,Go19,Ki20a,Sh20,Zh20b,Gh20,Ch21b,Ni21,To21a,Zh21,Yu21}. 

Torquato \cite{To21d} showed that in instances in which the spectral density has the power-law form 
\begin{equation}
{\tilde \chi}_{_V}({\bf k})\sim B|{\bf k}|^\alpha
\label{tildechi_alpha}
\end{equation}
in the limit $|\bf k|\rightarrow 0$, the long-time excess spreadability for two-phase media in $\mathbb{R}^d$ is given by the inverse power-law decay 
\begin{equation}
    {\cal S}(\infty)- {\cal S}(t)\sim \frac{B\Gamma[(d+\alpha)/2]\phi_2}{2^d\pi^{d/2}\Gamma(d/2)(Dt/a^2)^{(d+\alpha)/2}},
    \label{Spr_alpha}
\end{equation}
where $a$ represents some characteristic heterogeneity length scale. The power-law (\ref{Spr_alpha}) holds except when $\alpha\rightarrow +\infty$, which corresponds to stealthy hyperuniform media with a decay rate that is exponentially fast. The spreadability was computed for certain idealized ordered and disordered model microstructures across dimensions. Torquato \cite{To21d} also showed that $\mathcal{S}(t)$ has remarkable connections to covering problem of discrete geometry \cite{Co93,To10d} and nuclear magnetic resonance (NMR) or magnetic resonance imaging (MRI) measurements \cite{Mit92b,Se94,No14}. Specifically, Torquato \cite{To21d} identified precise mappings between the long-time formulas of the spreadability and the NMR pulsed field gradient spin-echo (PFGSE) amplitude \cite{Mit92b, Se94} as well as diffusion MRI measurements \cite{No14}. Thus, any analysis of the large-time behaviors of $\mathcal{S}(t)$ translates immediately into corresponding analyses of NMR or MRI data, which enables one to extract large-scale microstructural information of real heterogeneous media from diffusion measurements alone.

\begin{figure*}[!ht]
    \centering\includegraphics[width=10cm]{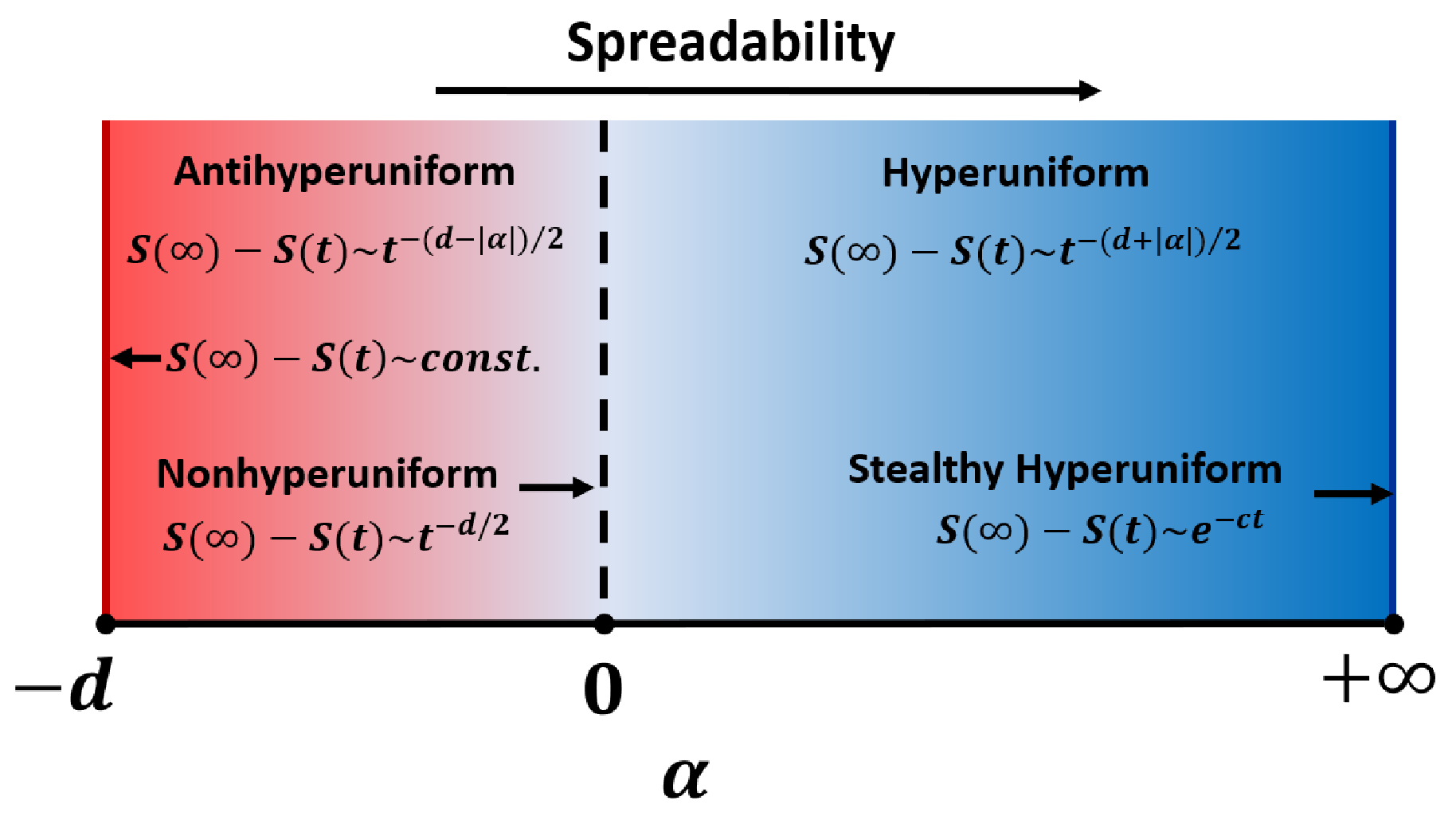}
  \caption{``Phase diagram'' that schematically shows the spectrum of long-time spreadability regimes in terms of the exponent $\alpha$, which depends on the large-scale properties of the microstructure. As $\alpha$ increases from the extreme antihyperuniform limit of $\alpha\rightarrow d$, the spreadability decay rate at long times decays faster. i.e., the excess spreadability follows the inverse power law $1/t^{(d+\alpha)/2}$, except when $\alpha\rightarrow\infty$, which corresponds to stealthy hyperuniform media with a decay rate that is exponentially fast. This figure is reproduced from the one presented in Ref. \cite{To21d}.}
  \label{spectrum}
\end{figure*}

\begin{figure*}[!ht]
\subfloat[]{
    \centering\includegraphics[width=3cm]{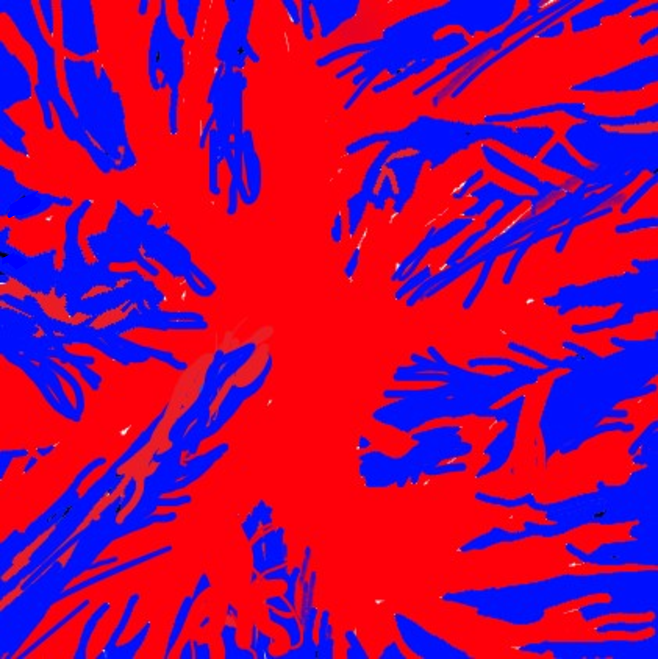}
}
\subfloat[]{
    \centering\includegraphics[width=3cm]{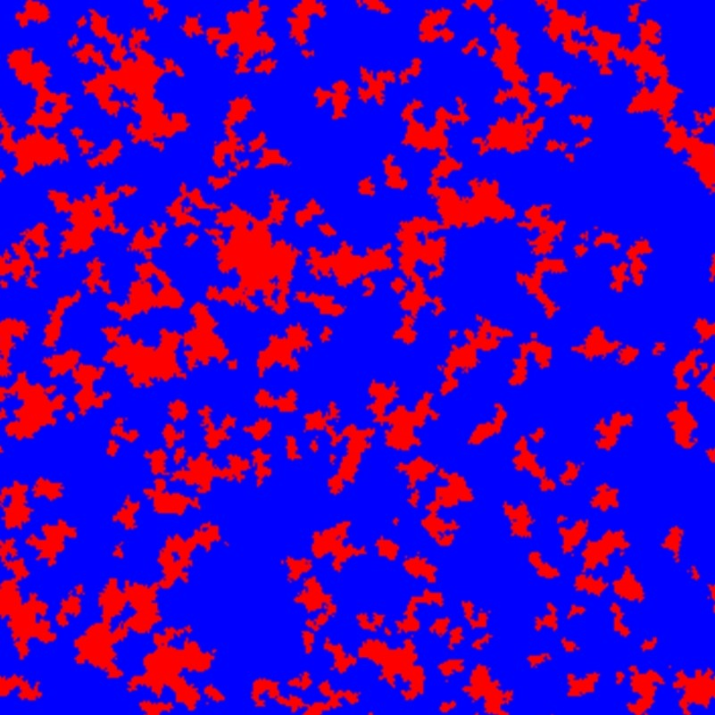}
}
\subfloat[]{
    \centering\includegraphics[width=3cm]{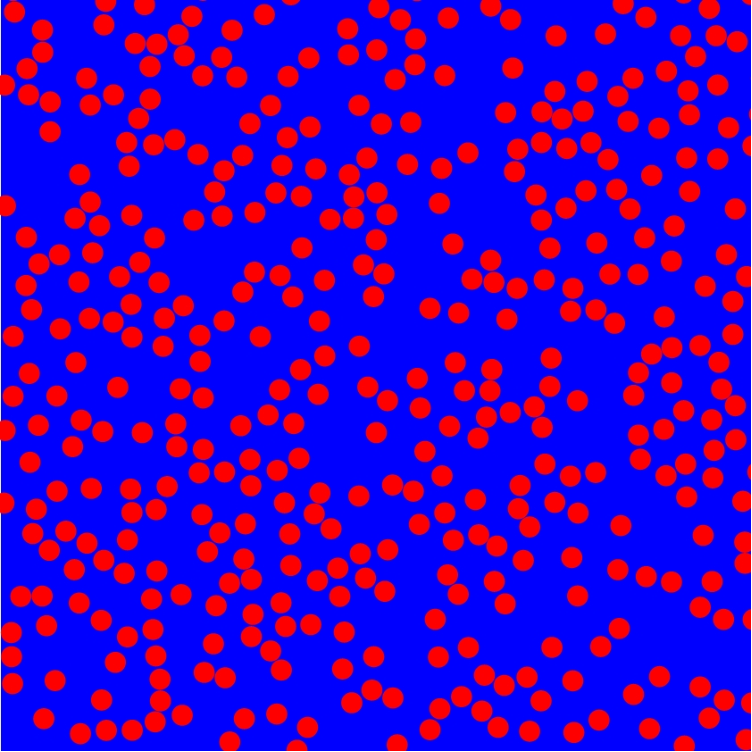}
}
\subfloat[]{
    \centering\includegraphics[width=3cm]{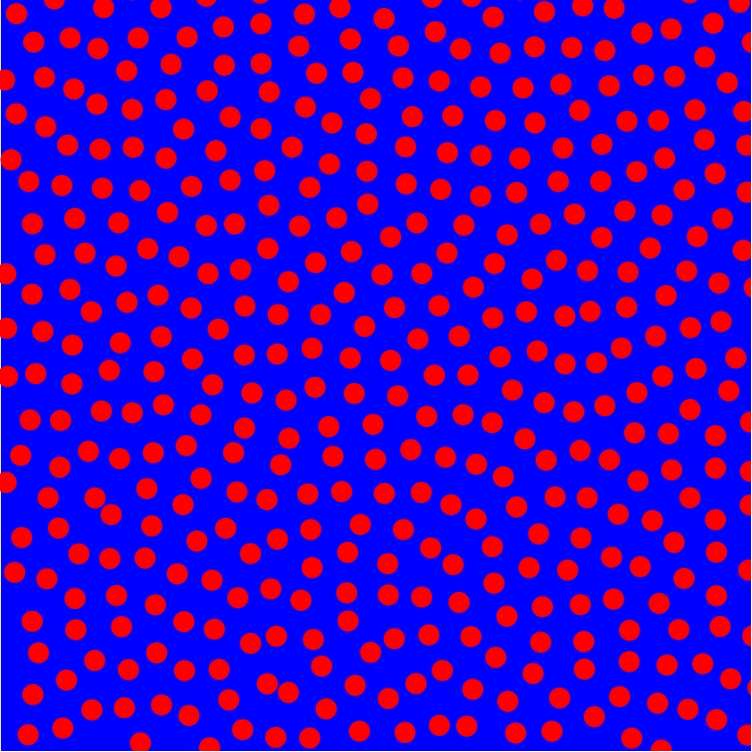}
}
\subfloat[]{
    \centering\includegraphics[width=3cm]{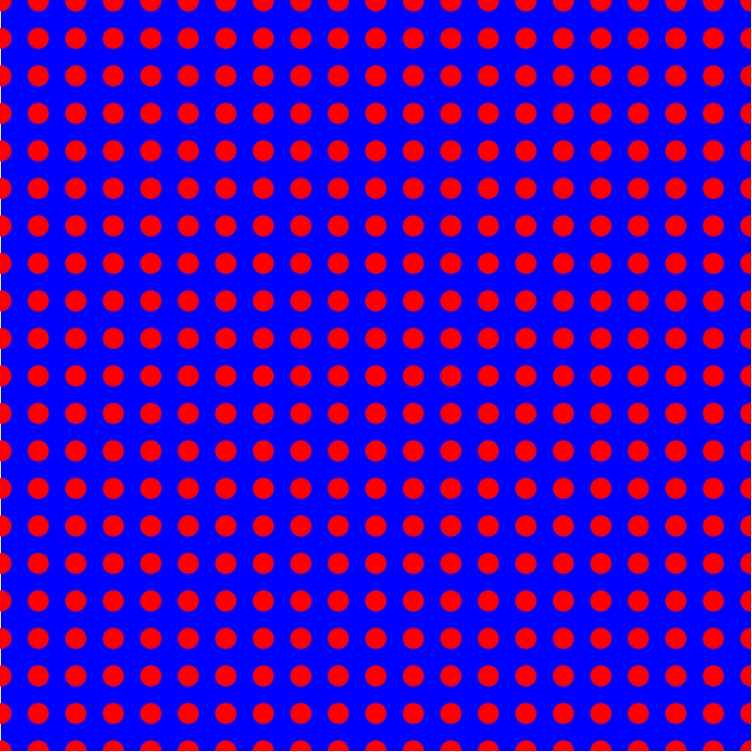}
}
  \caption{Specific examples of microstructures corresponding to the generic ones indicated in Fig. \ref{spectrum}. (a) Antihyperuniform media ($\alpha=-d$). (b) Typical disordered nonhyperuniform media ($\alpha=0$). (c) Disordered nonstealthy hyperuniform media ($0<\alpha<\infty$). (d) Disordered stealthy hyperuniform media ($\alpha=\infty$). (e) Ordered stealthy hyperuniform media ($\alpha=\infty$).}
  \label{cartoons}
\end{figure*}

Figure \ref{spectrum} schematically shows the ``spectrum'' of spreadability regimes in terms of the exponent $\alpha$. The long-time excess spreadability for antihyperuniform media ($-d < \alpha < 0$) have the slowest decay among all translationally invariant media, the slowest being when $\mathcal{S}(\infty)-\mathcal{S}(t)$ approaches a constant. A long-time decay rate of $\mathcal{S}(\infty) - \mathcal{S}(t) \sim t^{-d/2}$ corresponds to a nonhyperuniform medium ($\alpha=0$) in which the spectral density is a bounded positive number at the origin. The spreadability for nonstealthy hyperuniform media ($0<\alpha<\infty$) has the aforementioned power-law decay $\mathcal{S}(\infty) - \mathcal{S}(t) \sim t^{-(d+\alpha)/2}$. The limit $\alpha\rightarrow +\infty$ corresponds to media in which the decay rate of $\mathcal{S}$ is faster than any inverse power law, which is the case for stealthy hyperuniform media. Figure \ref{cartoons} shows specific examples of model microstructures corresponding to the generic ones indicated in Fig. \ref{spectrum}. The reader is referred to Sec. \ref{def} for terminology.

In this work, we further study the applicability of the spreadability concept as a dynamic measure to probe a wide range of hyperuniform and nonhyperuniform two-phase media. We explicitly compute spreadabilities of 2D and 3D models heretofore not previously considered that represent diverse hyperuniform and nonhyperuniform classes (see Sec. \ref{hyperuniformity}), including fully penetrable spheres \cite{To83b,To02a}, equilibrium hard spheres \cite{Wertheim1963,To02a,To18b}, as well as sphere packings derived from uniformly randomized lattices (URL) \cite{Klatt2020}, perfect glasses \cite{Zhang2016} and disordered stealthy point processes \cite{Zhang2016b,To18a}. These are compared and contrasted with spreadabilities for hyperuniform and nonhyperuniform media previously obtained by Torquato \cite{To21d}, including those for Debye random media \cite{De57,Ye98a} and sphere packings derived from Bravais lattices. We find that the small-time behavior of $\mathcal{S}(t)$ is determined by the derivatives of the autocovariance function $\chi_{_V}({\mathbf{r}})$ at the origin, the leading term of order $t^{1/2}$ being proportional to the specific surface $s$. The corresponding long-time behavior is determined by the form of the spectral density $\tilde{\chi}_{_V}(\mathbf{k})$ at small wavenumbers, which enables one to easily and robustly ascertain the class of hyperuniform or nonhyperuniform media. We also find that for URL packings, the excess spreadability has exponential decay (characteristic of the unperturbed lattice packing) at small to intermediate times, but transforms to power-law decay for $t$ larger than a well-defined transition time. Our study of the spreadability of the aforementioned models enables us to formulate an algorithm that efficiently and accurately extracts large-scale characteristics from spreadability data (or equivalently, NMR and MRI data) alone. Lessons learned from such analyses of our models is used to characterize the large-scale behaviors of a sample Fontainebleau sandstone \cite{Co96}, which we show is nonhyperuniform. Our study demonstrates the practical capability of the diffusion spreadability to extract crucial microstructural information of a class of heterogeneous media across length scales. We note that using reconstruction techniques that target autocovariance functions \cite{Ye98a,Za11f,Ma20b}, one could potentially construct two-phase media that realize prescribed functional forms of $\mathcal{S}(t)$, thereby designing materials with desired diffusion properties.

In Sec. \ref{def}, we provide basic definitions and preliminaries. We then describe the models that we consider in our study for $d=2,3$ and methods to compute their spreadabilities (Sec. \ref{models}). Section \ref{res} presents and discusses the spreadabilities of the aforementioned two-phase media models. Section \ref{algo} describes the algorithm for extracting large-scale structural characteristics from time-dependent diffusion behaviors, as measured by the spreadability, or equivalently, by NMR or MRI measurements. Section \ref{sec:fontainebleau} applies the lessons learned from the spreadabilities of the idealized models to analyze the structural characteristics of a sample Fontainebleau sandstone. Finally, Sec. \ref{conc} provides some conclusions and proposes potential future work. 

\section{Definitions and preliminaries}
\label{def}
In  this  section, we introduce some fundamental definitions and concepts that describe microstructures of two-phase media, including correlation functions, hyperuniformity, as well as the classification of hyperuniform and nonhyperuniform media. 

\subsection{Two-point statistics}

Consider a two-phase medium, i.e., a partition of space into two disjoint regions of phase 1 with volume fraction $\phi_1$ and of phase 2 of volume fraction $\phi_2$. For a statistically homogeneous medium, the {\it autocovariance} function $\chi_{_V}({\bf r})$ is directly related to the two-point correlation function $S_2^{(i)}({\bf r})$ of phase $i$ \cite{To02a}:
\begin{equation}
\chi_{_V}({\bf r}) \equiv S_2^{(1)}({\bf r})-\phi_1^2=S_2^{(2)}({\bf r})-\phi_2^2.
\label{covariance}
\end{equation}
The two-point function $S_2^{(i)}({\bf r})$ gives the probability that the two end points with displacement vector $\mathbf{r}$ are in phase $i$. At the extreme limits of its argument, $\chi_{_V}({\bf r})$ has the following asymptotic behavior: $\chi_{_V}({\bf r}=0)=\phi_1\phi_2$ and  $\lim_{|{\bf r}| \rightarrow \infty} \chi_{_V}({\bf r})=0$ if the medium possesses no long-range order. If the medium is statistically homogeneous and isotropic, then the autocovariance function ${\chi_{_V}}({\bf r})$ depends only on the magnitude of its argument $r=|\bf r|$, and hence is a radial function. In such instances, its slope at the origin is directly related to the {\it specific surface} $s$, which is the interface area per unit volume. In particular, the well-known three-dimensional asymptotic result \cite{De57} is easily obtained in any space dimension $d$:
\begin{equation}
\chi_{_V}({\bf r})= \phi_1\phi_2 - \kappa(d) s \;|{\bf r}| + {\cal O}(|{\bf r}|^2),
\label{specific}
\end{equation}
where
\begin{equation}
\kappa(d)= \frac{\Gamma(d/2)}{2\sqrt{\pi} \Gamma((d+1)/2)}.
\label{kappa}
\end{equation}
The nonnegative spectral density ${\tilde \chi}_{_V}({\bf k})$, which can be obtained from  scattering experiments \cite{De57}, is the Fourier transform of $\chi_{_V}({\bf r})$ at wave vector $\bf k$, i.e.,
\begin{equation}
{\tilde \chi}_{_V}({\bf k}) = \int_{\mathbb{R}^d} \chi_{_V}({\bf r}) e^{-i{\bf k \cdot r}} {\rm d} {\bf r} \ge 0, \qquad \mbox{for all} \; {\bf k}.
\label{spectral}
\end{equation}

A particular class of two-phase media is \textit{sphere packings}, i.e. a collection of spheres in $d$-dimensional Euclidean space $\mathbb{R}^d$ in which no two spheres overlap. 
The \textit{packing fraction} of a packing of identical spheres of radius $a$ is $\phi_2=\rho v_1(a)$, where $\rho$ is the number density and $v_1(a)$ is the volume of a sphere:
\begin{equation}
v_1(R)=\frac{\pi^{d/2} R^d}{\Gamma(1+d/2)}.
\label{v1}
\end{equation}
The spectral density of such a packing can be expressed as
\begin{equation}
{\tilde \chi}_{_V}({\bf k})  = \phi_2\, {\tilde \alpha}_2(k;a) S({\bf k}),
\label{chi-S}
\end{equation}
where $\tilde{\alpha}_2(k;a)$ is the Fourier transform of the scaled intersection volume of two spherical windows, which is given by \cite{To03a}
\begin{equation}
{\tilde \alpha}_2(k;a) = \Gamma(d/2+1)\frac{J_{d/2}^2(ka)}{k^d},
\end{equation}
and the structure factor $S(\mathbf{k})$ is defined as
	\begin{equation}
		S(\mathbf{k})=1+\rho \tilde{h}(\mathbf{k}),
		\label{skdef}
	\end{equation}
where $h(\mathbf{r})$ is the total correlation function, and $\tilde{h}(\mathbf{k})$ is the Fourier transform of $h(\mathbf{r})$.

For a single periodic configuration containing number $N$ point particles at positions ${\bf r}_1,{\bf r}_2,\ldots,{\bf r}_N$ within a fundamental cell $F$ of a lattice $\Lambda$, the {\it scattering intensity}  $I(\mathbf{k})$ is defined as
	\begin{equation}
		I(\mathbf{k})=\frac{\left|\sum_{i=1}^{N}e^{-i\mathbf{k}\cdot\mathbf{r}_i}\right|^2}{N}.
		\label{scattering}
	\end{equation}
For an ensemble of periodic configurations of $N$ particles within the fundamental cell $F$, the ensemble average of the scattering intensity in the infinite-volume limit is directly related to structure factor $S(\mathbf{k})$ by
	\begin{equation}
		\lim_{N,V_F\rightarrow\infty}\langle I(\mathbf{k})\rangle=(2\pi)^d\rho\delta(\mathbf{k})+S(\mathbf{k}),
	\end{equation}
where $V_F$ is the volume of the fundamental cell and $\delta(\mathbf{x})$ is the Dirac delta function  \cite{To18a}. In simulations of many-body systems with finite $N$ under periodic boundary conditions, Eq. (\ref{scattering}) is used to compute $S(\mathbf{k})$ directly by averaging over configurations.

\subsection{Hyperuniformity}
\label{hyperuniformity}
The \textit{hyperuniformity} concept generalizes the traditional notion of long-range order in many-particle systems to not only include all perfect crystals and perfect quasicrystals, but also exotic amorphous states of matter \cite{To03a,To18a}.
For two-phase heterogeneous media in $d$-dimensional Euclidean space $\mathbb{R}^d$, which include cellular solids, composites, and porous media, hyperuniformity is defined by the following infinite-wavelength  condition on the {\it spectral density} ${\tilde \chi}_{_V}({\bf k})$ \cite{Za09, To18a}, i.e.,
\begin{equation}
\lim_{|{\bf k}|\to 0 }{\tilde \chi}_{_V}({\bf k}) = 0.
\label{condition}
\end{equation}
An equivalent definition of hyperuniformity is based on the local volume-fraction
variance $\sigma^2_{_V}(R)$  associated with a $d$-dimensional spherical observation window  of radius $R$.
A two-phase medium in $\mathbb{R}^d$ is hyperuniform if its variance
grows in the large-$R$ limit faster than $R^d$.
This behavior is to be   contrasted with those of  typical disordered two-phase media for which the variance decays  like
the inverse of the volume $v_1(R)$ of the spherical observation window given by (\ref{v1}).
The  hyperuniformity  condition  (\ref{condition})  dictates  that  the  direct-space 
autocovariance function $\chi_{_V}({\bf r})$  exhibits both positive  and  negative  correlations  such  that  its  
volume integral over all space is exactly zero  \cite{To16b,To18a}, i.e.,
\begin{equation}
\int_{\mathbb{R}^d} \chi_{_V}({\bf r}) d{\bf r}=0,
\label{sum-rule}
\end{equation}
which is a direct-space sum rule for hyperuniformity.

\subsection{Classification of hyperuniform and nonhyperuniform media}
\label{class}
The hyperuniformity concept has led to a unified means to classify equilibrium and nonequilibrium states of matter, whether hyperuniform or not, according to their large-scale fluctuation characteristics. In the case of hyperuniform two-phase media  \cite{Za09,To18a},  there are three different scaling regimes (classes) that describe the associated large-$R$ behaviors of the volume-fraction variance when the spectral density goes to zero as a power-law scaling (\ref{tildechi_alpha}) in the limit $|\bf k|\rightarrow 0$ \cite{To18a}:
\begin{align}  
\sigma^2_{_V}(R) \sim 
\begin{cases}
R^{-(d+1)}, \quad\quad\quad \alpha >1 \qquad &\text{(Class I)}\\
R^{-(d+1)} \ln R, \quad \alpha = 1 \qquad &\text{(Class II)}\\
R^{-(d+\alpha)}, \quad 0 < \alpha < 1\qquad  &\text{(Class III).}
\end{cases}
\label{eq:classes}
\end{align}
Classes I and III are the strongest and weakest forms of hyperuniformity, respectively.
Class I media include all crystal structures, many quasicrystal structures and exotic
disordered media \cite{Za09,To18a}. Stealthy hyperuniform  media are also of class I and are defined to be those that possess 
zero-scattering intensity for a set of wavevectors around the origin \cite{To16b}, i.e.,
\begin{equation}
{\tilde \chi}_{_V}({\bf k})=0 \qquad \mbox{for}\; 0 \le |{\bf k}| \le K.
\label{stealth}
\end{equation}
Examples of such media are periodic packings of spheres
as well as unusual disordered sphere packings derived from stealthy point patterns \cite{To16b,Zhang2016b}.

By contrast, for any nonhyperuniform two-phase system, the local variance has the following large-$R$ scaling behaviors \cite{To21c}:
\begin{align}  
\begin{split}
\sigma^2_{_V}(R) \sim \begin{cases}
R^{-d}, &\alpha =0 \, \text{(typical nonhyperuniform)}\\
R^{-(d+\alpha)}, &-d <\alpha < 0 \, \text{ (antihyperuniform)}.\\
\end{cases}
\end{split}
\label{sigma-nonhyper}
\end{align}

For a  ``typical" nonhyperuniform system, ${\tilde \chi}_{_V}(0)$ is positive and bounded, and one has $B=\tilde{\chi}_{_V}(0)$ due to Eq. (\ref{tildechi_alpha}) \cite{To18a}. In {\it antihyperuniform} systems,
${\tilde \chi}_{_V}(0)$ is unbounded \cite{To21c}, i.e.,
\begin{equation}
\lim_{|{\bf k}| \to 0} {\tilde \chi}_{_V}({\bf k})=+\infty,
\label{antihyper}
\end{equation}
and hence  are diametrically opposite to hyperuniform systems. Antihyperuniform systems include  systems at thermal critical points (e.g., liquid-vapor and magnetic critical points) \cite{St87b,Bi92}, fractals \cite{Ma82}, disordered non-fractals \cite{To21b},
and certain substitution tilings \cite{Og19}.

\section{Two-phase media models}
\label{models}
\begin{figure*}[!ht]
\subfloat[]{
    \centering\includegraphics[width=4cm]{fig2b3a.eps}
}
\subfloat[]{
    \centering\includegraphics[width=4cm]{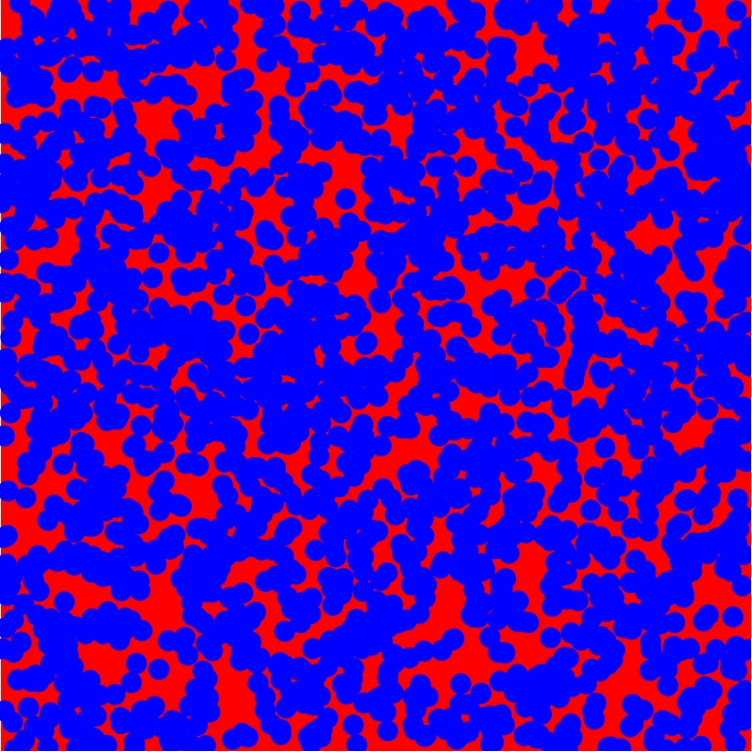}
}
\subfloat[]{
    \centering\includegraphics[width=4cm]{fig2c3c.eps}
}
\subfloat[]{
    \centering\includegraphics[width=4cm]{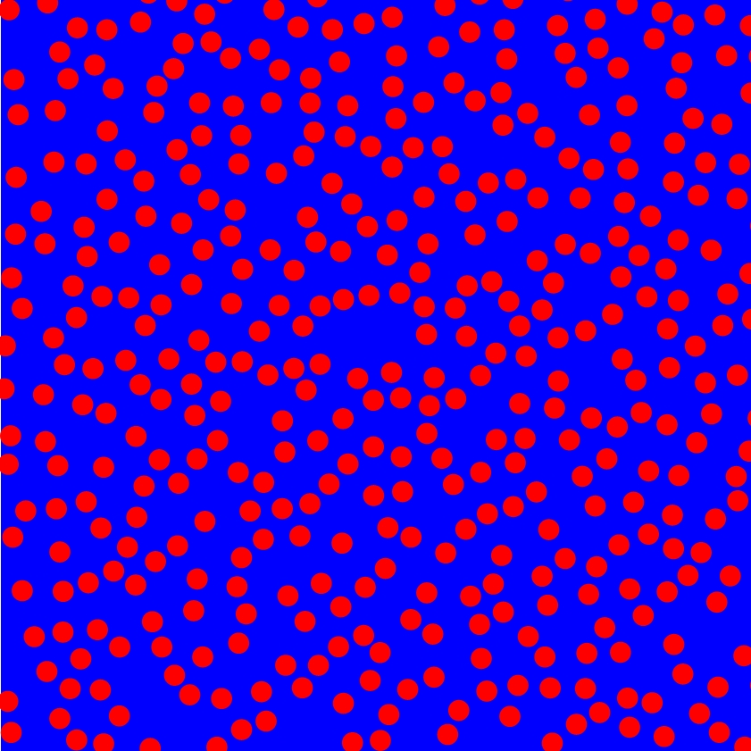}
}

\subfloat[]{
    \centering\includegraphics[width=4cm]{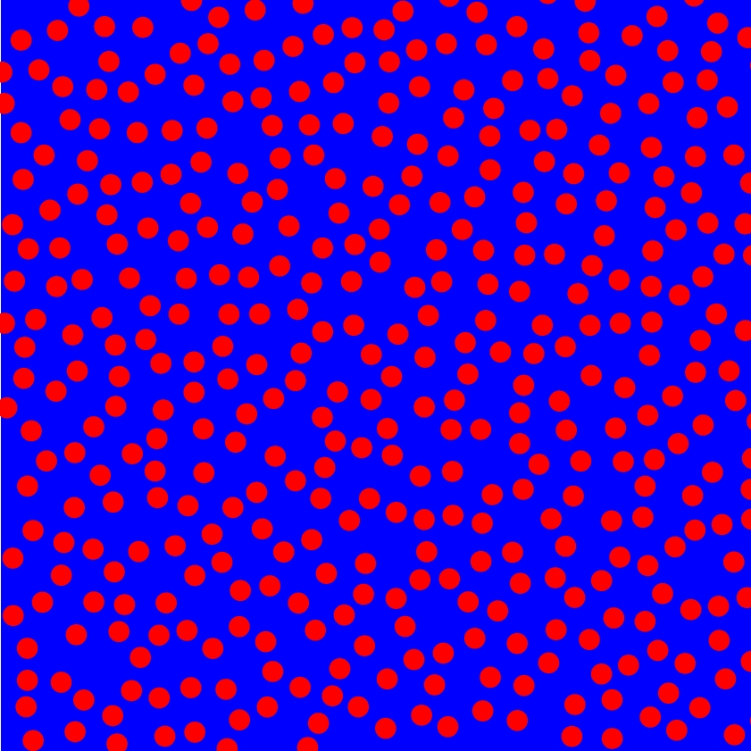}
}
\subfloat[]{
    \centering\includegraphics[width=4cm]{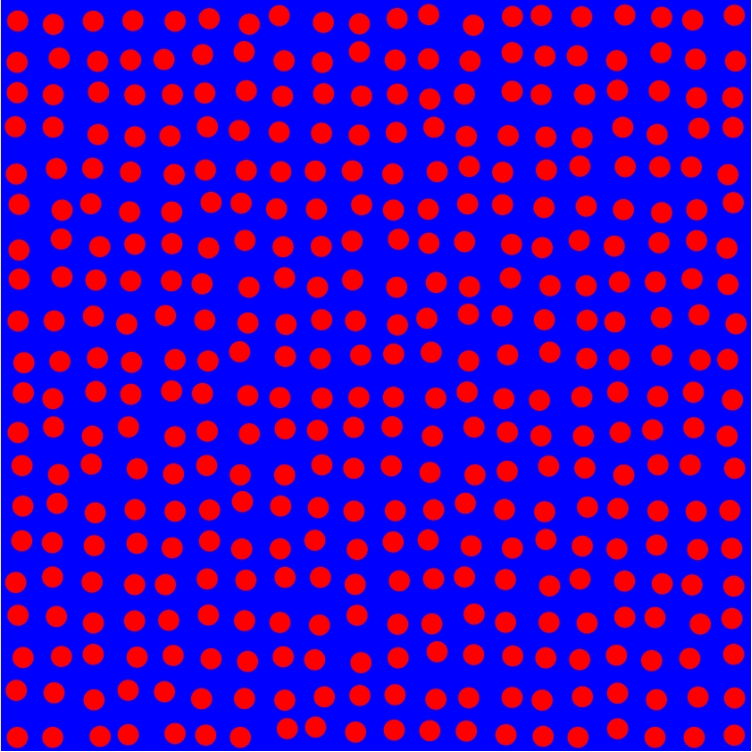}
}
\subfloat[]{
    \centering\includegraphics[width=4cm]{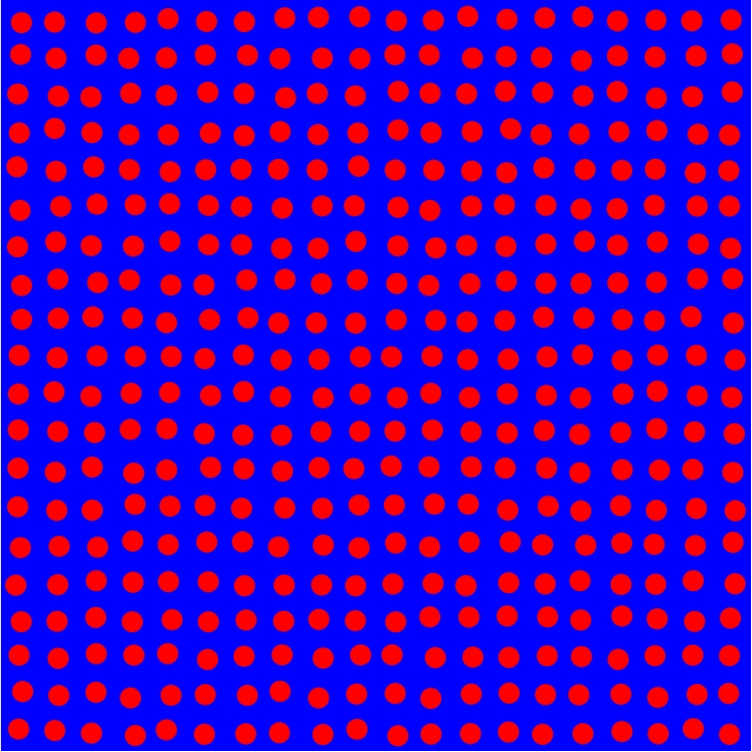}
}
\subfloat[]{
    \centering\includegraphics[width=4cm]{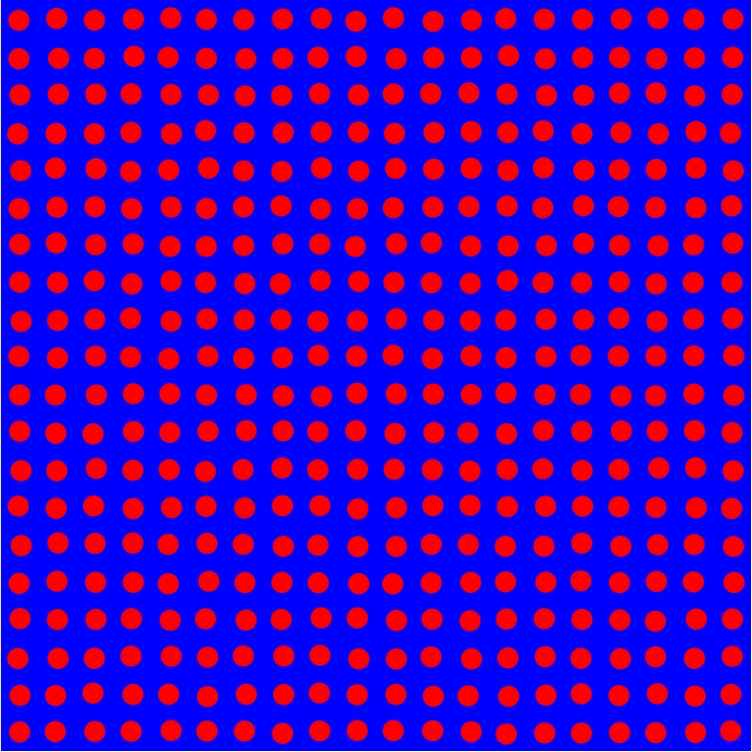}
}

\subfloat[]{
    \centering\includegraphics[width=4cm]{fig2d3i.eps}
}
\subfloat[]{
    \centering\includegraphics[width=4cm]{fig2e3j.eps}
}
\subfloat[]{
    \centering\includegraphics[width=4cm]{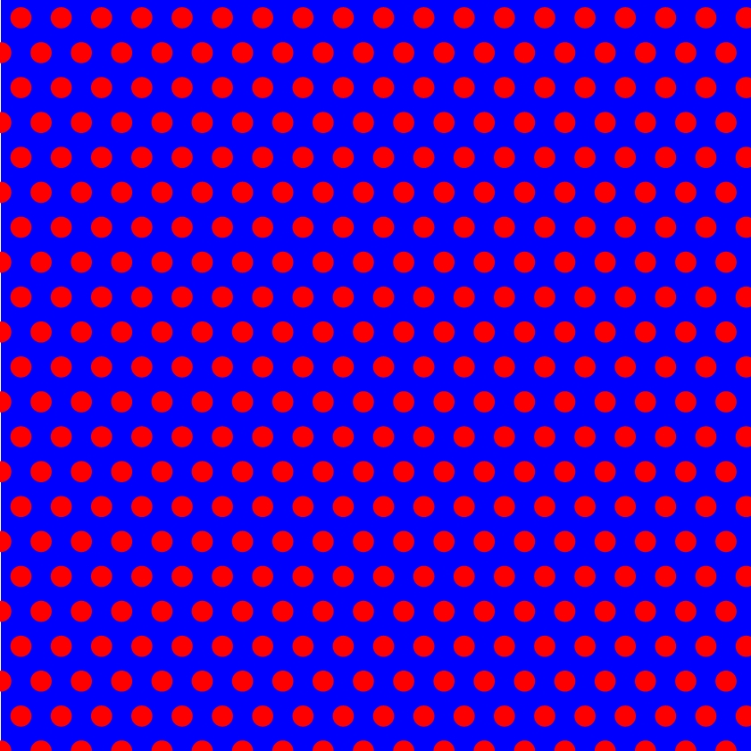}
}
  \caption{Two-phase media in two dimensions with $\phi_2=0.25$ studied in this work. A square portion of side length $70.90a$ is shown for each medium. Phase 2 is colored red. (a) Debye random media. (b) Fully penetrable circular disks. (c) Equilibrium hard disks. (d) Perfect glass disk packing with $\alpha=1$. (e) Perfect glass disk packing with $\alpha=2$. (f) Disk packing corresponding to URL with $b=0.3$. (g) Disk packing corresponding to URL with $b=0.2$. (h) Sphere packing corresponding to URL with $b=0.1$. (i) Disordered stealthy hyperuniform disk packing with $Ka=1.3$. (j) Square-lattice disk packing. (k) Triangle-lattice disk packing.}
  \label{2dSnaps}
\end{figure*}

\begin{figure*}[htp]
\subfloat[]{
    \centering\includegraphics[width=4cm]{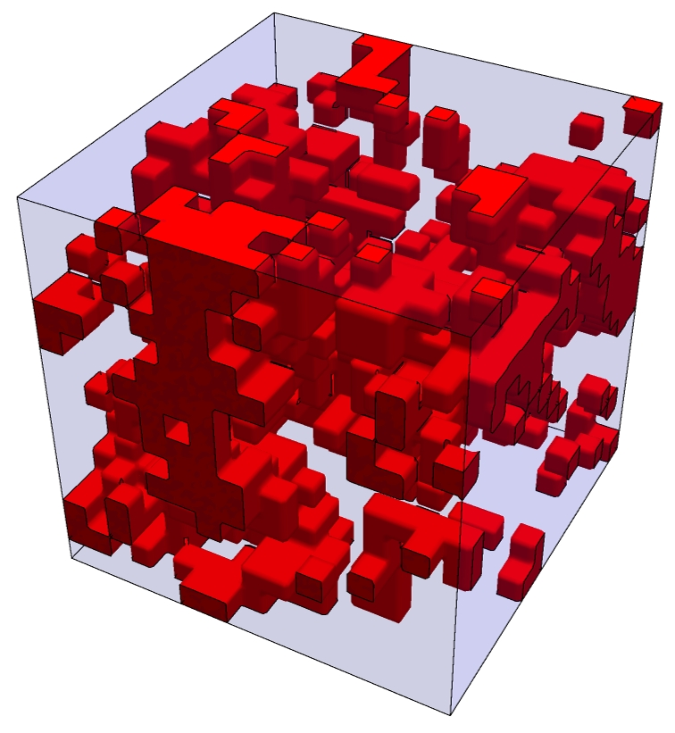}
}
\subfloat[]{
    \centering\includegraphics[width=4cm]{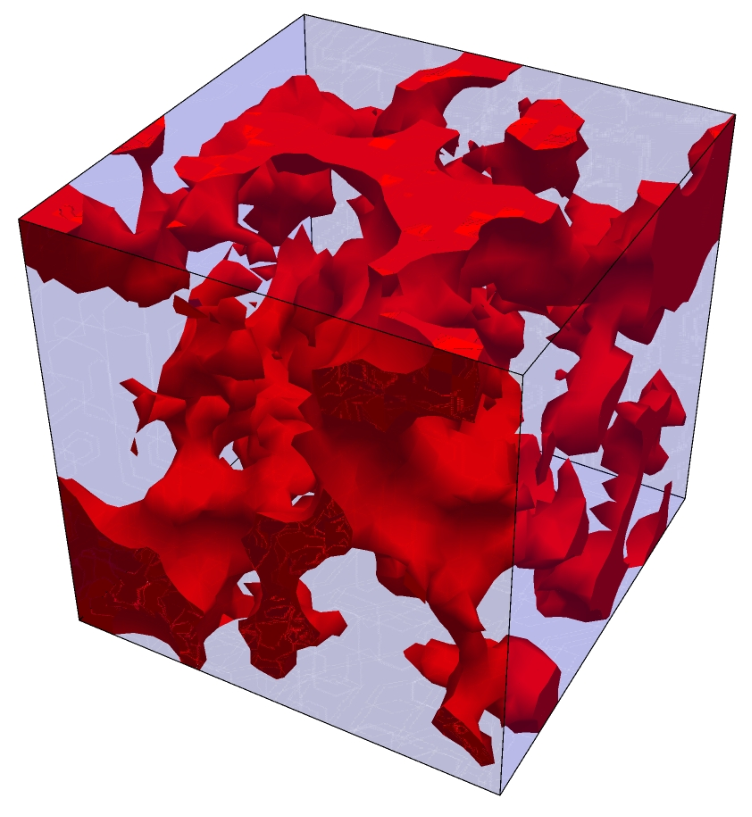}
}
\subfloat[]{
    \centering\includegraphics[width=4cm]{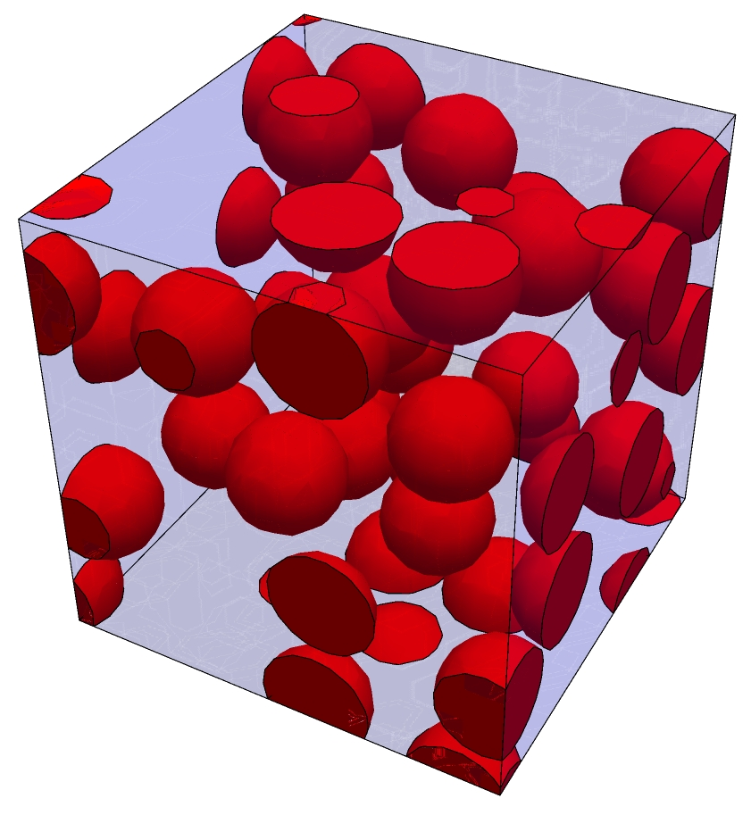}
}
\subfloat[]{
    \centering\includegraphics[width=4cm]{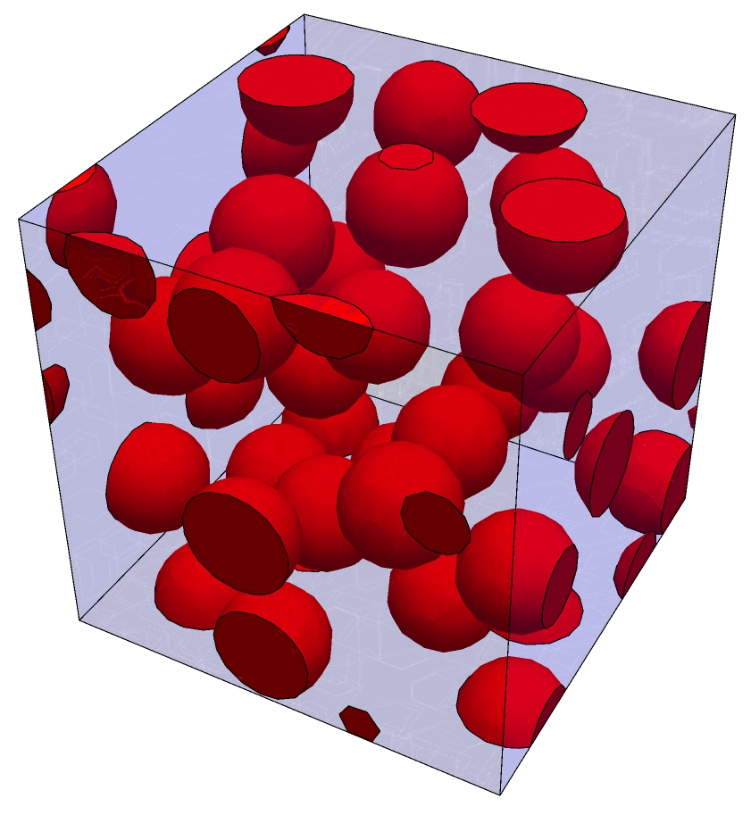}
}

\subfloat[]{
    \centering\includegraphics[width=4cm]{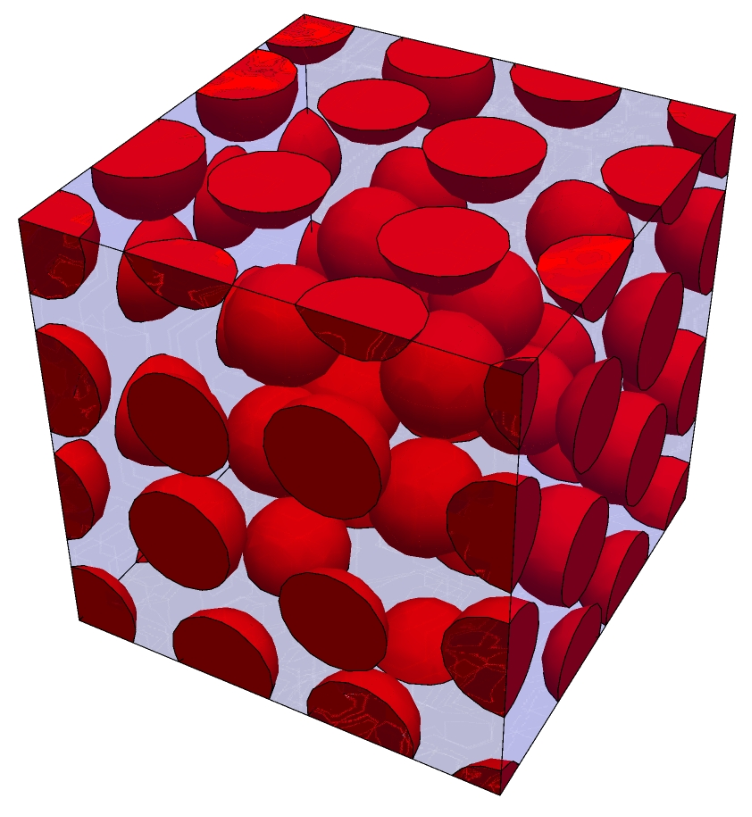}
}
\subfloat[]{
    \centering\includegraphics[width=4cm]{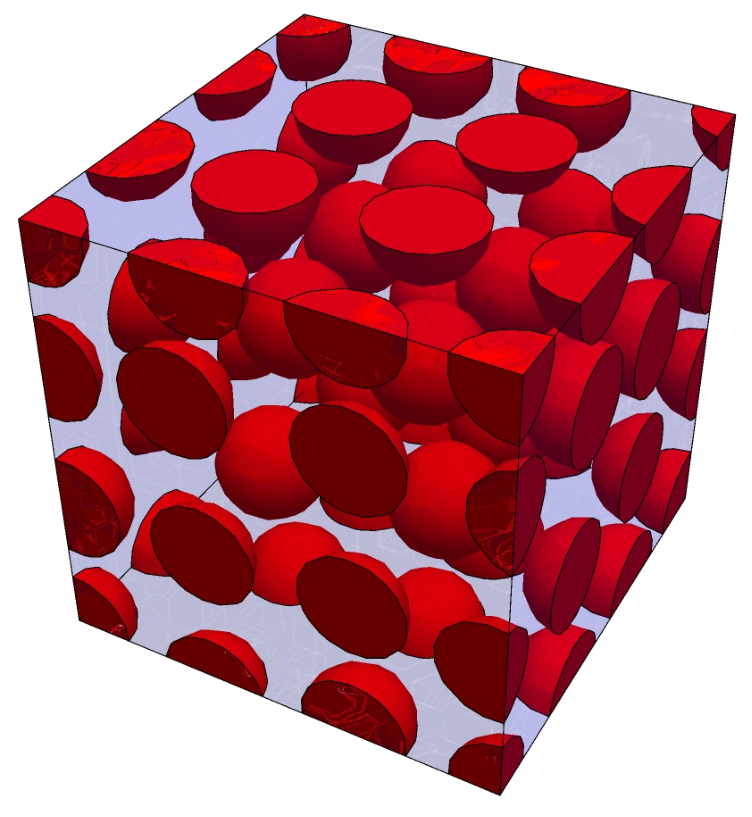}
}
\subfloat[]{
    \centering\includegraphics[width=4cm]{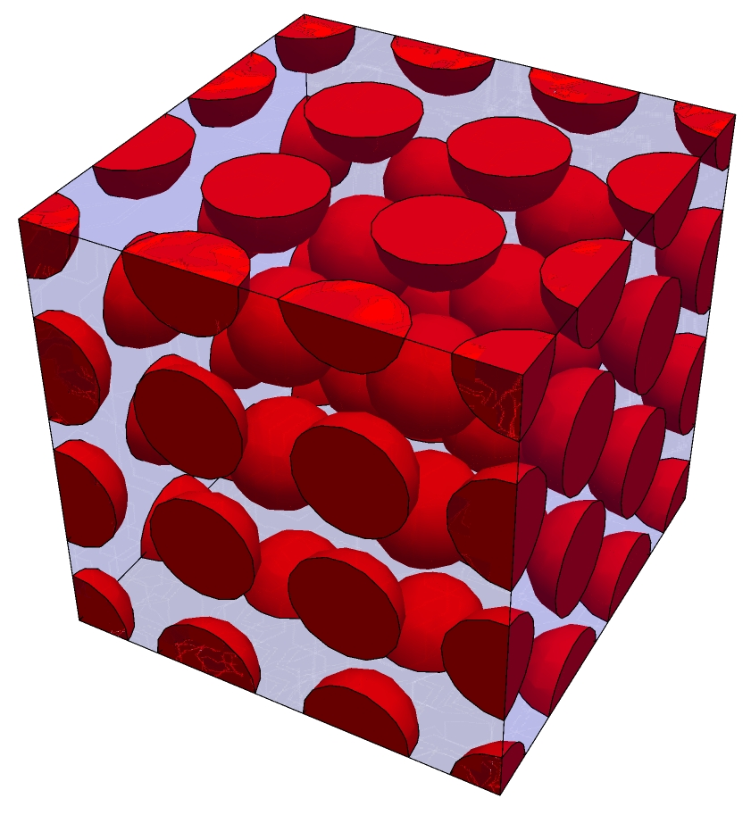}
}
\subfloat[]{
    \centering\includegraphics[width=4cm]{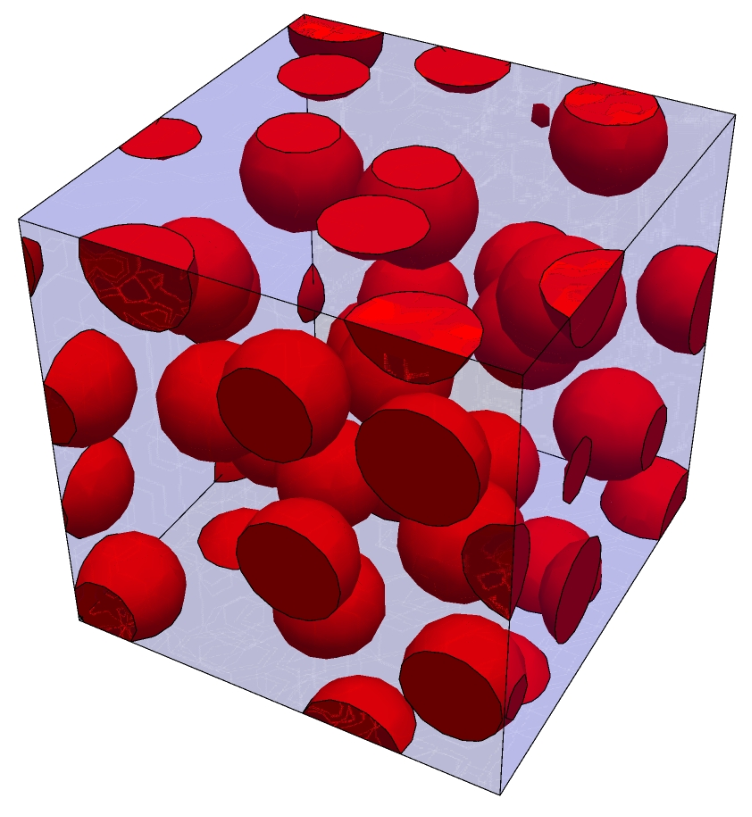}
}

\subfloat[]{
    \centering\includegraphics[width=4cm]{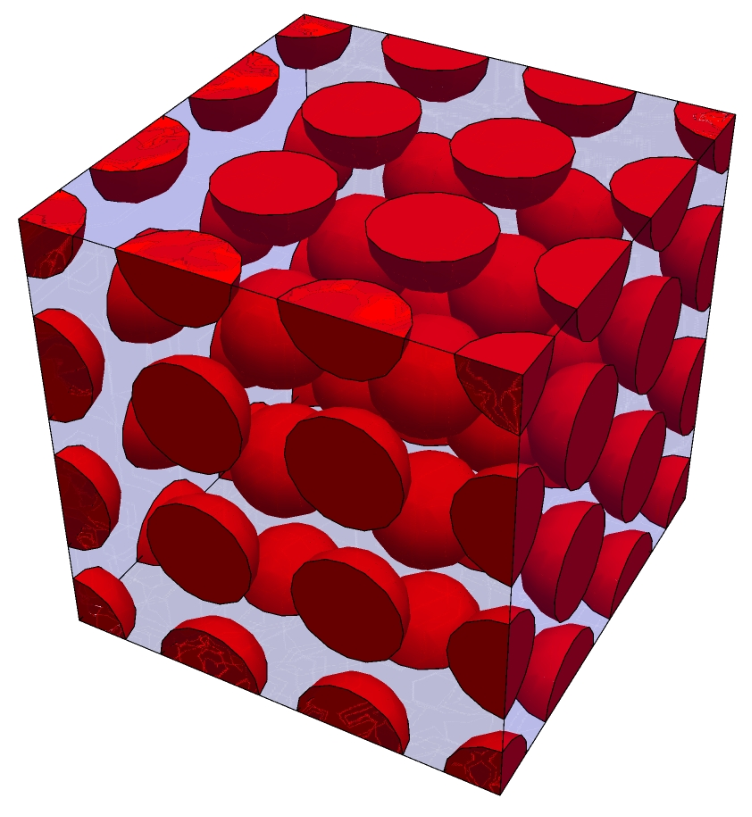}
}
\subfloat[]{
    \centering\includegraphics[width=4cm]{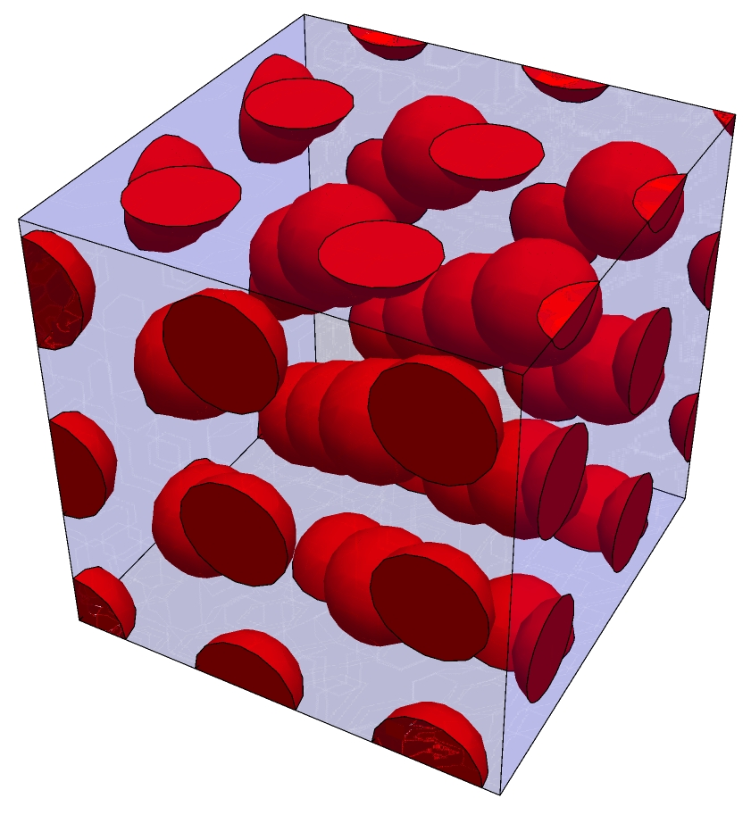}
}
  \caption{Two-phase media in three dimensions with $\phi_2=0.20$ studied in this work. Unless otherwise stated, a cubic portion of side length $8.21a$ is shown for each model. Phase 2 is colored red. (a) Debye random media, a cubic portion of side length $16a$ is shown here to better represent the distribution of the phases. (b) Fully penetrable spheres. (c) Equilibrium hard spheres. (d) Perfect glass sphere packing with $\alpha=1$. (e) Sphere packing corresponding to URL with $b=0.3$. (f) Sphere packing corresponding to URL with $b=0.2$. (g) Sphere packing corresponding to URL with $b=0.1$. (h) Disordered stealthy hyperuniform sphere packing with $Ka=1.5$. (i) Cubic-lattice sphere packing. (j) BCC-lattice sphere packing. }
  \label{3dSnaps}
\end{figure*}

To further explore the potential of the spreadability as a dynamic-based figure of merit to probe microstructures of two-phase media, we compute it for a variety of models in two and three space dimensions that represent diverse hyperuniform and nonhyperuniform classes, as described in Sec. \ref{class}. The nonhyperuniform models that we study include fully penetrable spheres  \cite{To83b,To02a},  equilibrium hard spheres \cite{To02a,Weeks1971} and Debye random media  \cite{De57,Ye98a}. The hyperuniform models that we study include monodisperse sphere packings derived from perfect glasses \cite{Zhang2016}, uniformly randomized lattices (URL) \cite{Klatt2020}, disordered stealthy hyperuniform point processes \cite{Zhang2016b} and Bravais lattices. We describe each model below and the methods to compute their spreadabilities. We choose the volume fraction $\phi_2$ of phase 2, (i.e., the phase initially containing the solute,) to be 0.25 for 2D models and 0.20 for 3D models. These relatively small values of $\phi_2$ were chosen so that we could derive sphere packings from the aforementioned point processes by decorating each point with a circular disk for $d=2$ or a sphere for $d=3$. We also note that for our chosen $\phi_2$'s, the values of $\phi_2/\phi_{m,d}$ are similar for $d=2,3$, where $\phi_{m,2}=0.9069$ and $\phi_{m,3}=0.7405$ are the maximum packing fractions for sphere packings in $\mathbb{R}^2$ and $\mathbb{R}^3$, respectively. This facilitates comparison of spreadabilities across dimensions. For all packings, phase 2 is chosen to be the particle phase. Figures \ref{2dSnaps} and \ref{3dSnaps} show snapshots of all the models considered in this study.

\subsection{Nonhyperuniform fully penetrable spheres}
We consider the well-known model
of ``fully penetrable spheres'' (FPS) \cite{To83b,To02a}, which is nonhyperuniform. It is constructed by decorating each point of a Poisson point process by a $d$-dimensional sphere of radius $a$ that generally may overlap with one another. The simplicity of the $n$-body statistics of the Poisson process enables us to exactly express $\chi_{_V}(r)$ of the \textit{matrix} phase, i.e., the space unoccupied by the particles, which is designated phase 2 in this work \cite{To02a}:
\begin{equation}
    \chi_{_V}(r) = \exp\left[-\eta\frac{v_2(r;a)}{v_1(a)} \right]-\phi_2^2,
    \label{chi_fps}
\end{equation}
where $\eta=-\ln\left(\phi_2\right)$ and $v_2(r;a)$ is the union volume of two $d$-dimensional spheres of radius $a$. Specifically,
\begin{widetext}
\begin{equation}
    \frac{v_2(r;a)}{v_1(a)} =2\Theta(r-2a)
    +\frac{2}{\pi}\left[\pi + \frac{r}{2a}\left(1-\frac{r^2}{4a^2}\right)^{\frac{1}{2}} -\cos^{-1}\left(\frac{r}{2a}\right)\right]\Theta\left( 2a-r \right), \qquad d=2,
\end{equation}
\begin{equation}
    \frac{v_2(r;a)}{v_1(a)}=2\Theta(r-2a)+\left[1+\frac{3r}{4a}-\frac{r^3}{16a^3}\right]\Theta\left( 2a-r \right), 
    \qquad d=3.
\end{equation}
\end{widetext}
The specific surface for fully penetrable spheres is given by \cite{To84a}
\begin{equation}
   as = -d\phi_2\ln(\phi_2).
   \label{fps_s}
\end{equation}

\subsection{Nonhyperuniform equilibrium hard spheres}
Equilibrium hard sphere packings provide good models of condensed matter in both liquid and crystal states when the interactions are dominated
by strong short-range repulsions \cite{Weeks1971,Cheng1999}. We study such systems along the stable fluid branch \cite{To02a,To18a}. We generated ensembles of 1,000 equilibrium hard-sphere configurations using Monte Carlo simulations in 2D and 3D at the aforementioned values of $\phi_2$. The simulations used square or cubic simulation boxes under periodic boundary conditions. The ensemble-averaged structure factors were then computed from Eq. (\ref{scattering}). The specific surface for all sphere packings is given by \cite{To84a}
\begin{equation}
   as = d\phi_2.
   \label{packing_s}
\end{equation}

\subsection{Nonhyperuniform Debye random media}
Useful models of nonhyperuniform two-phase media are Debye random media \cite{De57,Ye98a}, defined entirely by the following monotonic radial autocovariance function:
\begin{equation}
\chi_{_V}(r) =\phi_1\phi_2 \exp(-r/a),
\label{debye}
\end{equation}
where $a>0$ is a length-scale parameter. Such media can never be hyperuniform because the sum rule (\ref{sum-rule}) requires both positive and negative correlations \cite{To16b}. Debye et al. \cite{De57} hypothesized the simple exponential form (\ref{debye})  to model three-dimensional media with phases of ``fully random shape, size, and distribution.” Such autocovariance functions were shown to be realizable in two \cite{Ye98a,Chi13,Ma20b} and three \cite{Ji07,To20a} dimensions. Torquato obtained explicit expressions of $\mathcal{S}(t)$ for Debye random media in any space dimension \cite{To21d}. We consider them in this study as they are good approximation of certain realistic heterogeneous materials \cite{De57}, including Fontainebleau sandstones \cite{Co96}. They also serve as ``benchmark'' models to test our algorithm that extracts large-scale microstructural information from time-dependent diffusion data, as their spreadabilities are exactly known. The specific surface for $d$-dimensional Debye random media is given by \cite{To21d}
\begin{equation}
    as = \frac{\phi_1\phi_2}{\kappa(d)},
    \label{debye_s}
\end{equation}
where $\kappa(d)$ is given by (\ref{kappa}).

\subsection{Hyperuniform sphere packings derived from ``perfect glasses''}
\textit{Perfect glasses} proposed by Zhang et al. \cite{Zhang2016} are hyperuniform many-body systems with positive bulk and shear moduli that banish any crystalline or quasicrystalline phases and are unique disordered states up to trivial symmetries \cite{Zh17b}. Thus, they form soft-interaction analogs of the maximally random jammed (MRJ) packings \cite{To00b} of hard particles. These latter states can be regarded as prototypical glasses since they are out of equilibrium, maximally disordered, hyperuniform, mechanically rigid with infinite bulk and shear moduli, and can never crystallize due to configuration-space trapping. A perfect glass
is created by cooling  many-body systems with a potential that involves certain two-, three-, and four-body soft interactions upon cooling from high temperature to zero temperature \cite{Zhang2016}. Using this procedure, Zhang et al. \cite{Zhang2016} generated perfect-glass point configurations with various values of $\alpha>0$ in both 2D and 3D. Specifically, their 2D and 3D perfect glasses with $\alpha=1$ and $\alpha=2$ can be transformed into packings with our prescribed packing fractions by decorating each point with a sphere of radius $a$  \cite{Zhang2016b}. We obtained numerical structure factors for these perfect glasses, which are available as supplementary material of Ref. 31. 

\subsection{Hyperuniform sphere packings derived from uniformly randomized lattices (URL) }
Perturbed lattices serve as important models in cosmology, crystallography and probability theory \cite{Welberry1980,Gabrielli2002}. Consider a perturbed lattice in which each point $\mathbf{x}$ of a lattice $\mathcal{L}$ is perturbed independently by a displacement vector $\mathbf{u}_\mathbf{x}$ that follows the probability density function $f(\mathbf{u_x})$. The structure factor $S(\mathbf{k})$ of this point process is given by \cite{Ga04,Ki18a}
\begin{equation}
    S(\mathbf{k})=1-|\tilde{f}(\mathbf{k})|^2+|\tilde{f}(\mathbf{k})|^2 S_{\mathcal{L}}(\mathbf{k}),
    \label{s_perturbedLatt}
\end{equation}
where $S_{\mathcal{L}}(\mathbf{k})$ is the structure factor of the unperturbed lattice $\mathcal{L}$ and $\tilde{f}(\mathbf{k})$ is the characteristic function, i.e., the Fourier transform of $f$. 

Here, we study a special class of hyperuniform sphere packings associated with \textit{uniformly randomized lattice} (URL) models derived from cubic lattices, introduced by Klatt et al. \cite{Klatt2020}. In these models, each point in a $d$-dimensional simple cubic lattice $\mathcal{L}=\mathbb{Z}^d$ is displaced by a random vector that is uniformly distributed on a rescaled unit cell $b C = [-b/2,b/2)^d$, where $b > 0$ is a scalar factor and $C$ is a unit cell of the lattice. Equation (\ref{s_perturbedLatt}) implies that URL point processes are Class I hyperuniform with $\alpha=2$. The \hyperref[sec:urlfTilde]{Appendix} gives explicit formulas for the characteristic function $\tilde{f}(\mathbf{k})$ of the uniform perturbation on $[-b/2,b/2)^d$, as well as plots of the angular averaged function $1-|\tilde{f}(k)|^2$, which is equal to the diffuse part of the structure factor $S(k)$ and can be numerically evaluated with arbitrary precision. It can be shown that in the limit $b\rightarrow 0$ and $k\rightarrow 0$, one has $1-|\tilde{f}(k)|^2 \propto b^2k^2$ \cite{Klatt2020}. Remarkably, Klatt et al. showed that the Bragg peaks of URL models disappear when $b$ takes integer values \cite{Klatt2020}. We derived monodisperse 2D and 3D sphere packings from URL models by decorating each point with a sphere of radius $a$, which is determined by the prescribed $\phi_2$ values. We verified that the spheres do not overlap for $b\leq 0.3$ in both 2D and 3D.

\subsection{Disordered stealthy hyperuniform sphere packings}
Disordered stealthy hyperuniform materials are exotic amorphous states of matter that have unusual structural characteristics (hidden order at large length scales) and physical properties, including desirable photonic and transport properties \cite{To16b,Zhang2016b}. Disordered stealthy hyperuniform packings have been generated using the collective-coordinate optimization procedure  \cite{To15,Di18} by decorating the resulting ground-state point configurations by nonoverlapping spheres of radius $a$ \cite{Zhang2016b,Ki20a}. The degree of order of such ground states depends on a tuning parameter $\chi$, which measures the extent to which the  ground  states  are  constrained by the size of the cut-off value $K$ relative to the number of degrees of freedom. For $\chi <1/2$, the ground states are typically disordered and uncountably infinitely degenerate in the infinite-volume limit  \cite{To15}. Using the fact that $\rho \chi=v_1(K)/[2d(2\pi)^d]$ \cite{To15}, it immediately follows that for identical nonoverlapping spheres of radius $a$ that the dimensionless stealthy cut-off value $Ka$ in terms of the packing fraction $\phi_2$ for any space dimension $d$ is given by
\begin{equation}
(Ka)^d=d 2^{d+1}\Gamma^2(1+d/2) \phi_2 \chi.
\label{K}
\end{equation}

We obtained numerical structure factors for disordered stealthy hyperuniform many-body systems with $N=1000, \rho=1$ in 2D and 3D described in Ref. \cite{Ki21}. The stealthy systems have $Ka=1.3$ for 2D and $Ka=1.5$ for 3D. 

\subsection{Ordered hyperuniform sphere packings}
For models of ordered hyperuniform media, we consider sphere packings derived from Bravais lattices in 2D and 3D, including the triangle lattice, the square lattice, the body-centered cubic (BCC) lattice and the cubic lattice. For each medium, the lattice points are decorated with identical spheres with radius $a$. The structure factor of the sites of a Bravais lattice in $\mathbb{R}^d$ is given by \cite{To21d}
\begin{equation}
S({\bf k}) =\frac{(2\pi)^d}{v_c} \sum_{{\bf Q} \neq {\bf 0}} \delta({\bf k} -{\bf Q}).
\label{factor}
\end{equation}
where $v_c$ is the volume of a fundamental cell in direct
space and $\mathbf{Q}$ denotes a reciprocal lattice (Bragg) vector.

\section{Results for the spreadabilities}
\label{res}

In this section, we present the results for the spreadabilities of the models in Sec. \ref{models}. We use the angular-averaged versions of Eq. (\ref{spreadability}) derived in Ref. \cite{To21d}, which were shown to be exact for translationally invariant two-phase media. To compare the spreadabilities of different models, we scale the microstructures such that all models possess unit specific surface, i.e, $s=1$. This simple microscopic length scaling choice has been previously applied by Kim and Torquato \cite{Ki21} because it has the advantage that $s$ is easily computable for many well-known models and $1/s$ is directly proportional to the arithmetic mean of the mean chord length $\ell^{(i)}_C$ of both phases \cite{To02a}.   Specifically, given the chosen values of $\phi_2$ in this study, Table \ref{table_a} lists the values of the characteristic dimensionless length scale $a^*=as$ for the models that we consider, obtained from (\ref{debye_s}), (\ref{fps_s}) and (\ref{packing_s}).

\begin{center}
\begin{table}
\caption{Values of the characteristic dimensionless length scale $a^*=as$ for the models in this study.}
\begin{tabular}{ ||c|c|| } 
\hline
 Model & $a^*=as$ \\
 \hline
 2D Debye & 0.589 \\ 
 2D FPS & 0.693 \\
 2D disk packings & 0.5 \\
 3D Debye & 0.64 \\
 3D FPS & 0.966 \\
 3D sphere packings & 0.6 \\
 \hline
\end{tabular}
\label{table_a}
\end{table}
\end{center}

It is instructive to compare our results for the spreadabilities for the aforementioned models to that for nonhyperuniform Debye random media \cite{De57,Ye98a,Ma20b,sk21}, as by Eq. (\ref{debye}). Torquato \cite{To21d} obtained closed-form expressions of the spreadability for a Debye random media in any space dimension $d$:
\begin{equation}
{\cal S}(\infty) -{\cal S}(t) =\frac{d\omega_d \phi_1}{(4\pi D t/a^2)^{d/2}} I_d(t),
\label{S-debye}
\end{equation}
where
\begin{equation}
I_d(t) =\frac{1}{a^d}\int_0^\infty r^{d-1} \exp(-r/a) \exp[-r^2/(4Dt)] dr.
\end{equation}
Specifically, in 2D and 3D,
\begin{widetext}
\begin{equation}
I_2(t)= \frac{2Dt}{a^2}\left\{1-\exp(Dt/a^2)\,\sqrt{\pi Dt/a^2}\,\left[1- \mbox{erf}(\sqrt{Dt/a^2})\right]\right\},
\end{equation}
\begin{equation}
I_3(t)= \frac{2Dt}{a^2}\Big\{ \exp(Dt/a^2)\,\sqrt{\pi Dt/a^2}\,\left[1 -\mbox{erf}(\sqrt{Dt/a^2})\right]\left[1+2Dt/a^2\right]- 2Dt/a^2\Big\}.
\end{equation}
\end{widetext}

The spreadabilities associated with fully penetrable spheres were computed from  (\ref{spreadability_direct}) and (\ref{chi_fps}). For this particular model, the direct-space representation (\ref{spreadability_direct}) is the preferred method to compute spreadability than the Fourier-space representation, as $\chi_{_V}(r)$ is known exactly and vanishes identically for $r\geq 2a$. Therefore, the integration in (\ref{spreadability_direct}) can be evaluated to arbitrary precision. On the other hand, the spreadabilities for all the sphere-packing models were computed by substituting (\ref{chi-S}) into the Fourier-space representation (\ref{spreadability_fourier}), because structure-factor data for sphere packings are readily available. For periodic sphere packings, Torquato \cite{To21d} obtained
\begin{equation}
{\cal S}(\infty)-{\cal S}(t)=\phi_2 \sum_{{\bf Q} \neq {\bf 0}}   \frac{{\tilde \alpha}_2(|{\bf Q}|;a)}{v_1(a)} \exp[-|{\bf Q}|^2 Dt].
\label{packing-b}
\end{equation}
In practice, we truncated the numerical integration in (\ref{spreadability_fourier}) at $|\mathbf{k}|a=100$. Similarly, the summation in (\ref{packing-b}) was truncated at $|\mathbf{Q}|a=100$. We verified that errors due to this truncation are negligible. 

\begin{figure*}[!ht]
\centering
\subfloat[]{
  \includegraphics[trim = 20 50 0 150, clip, width=8cm]{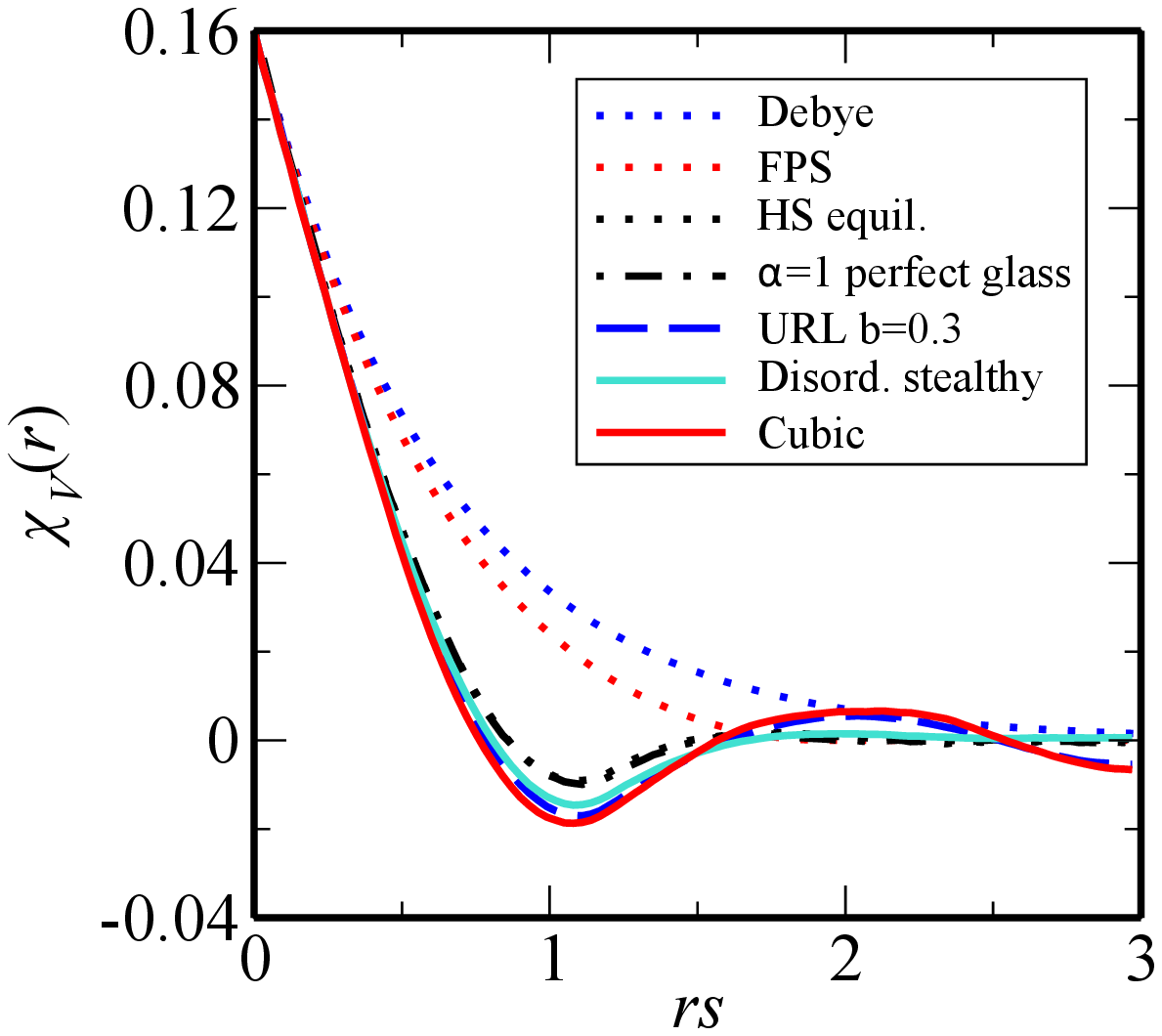}
}
\subfloat[]{
  \includegraphics[trim = 50 50 0 160, clip, width=7.5cm]{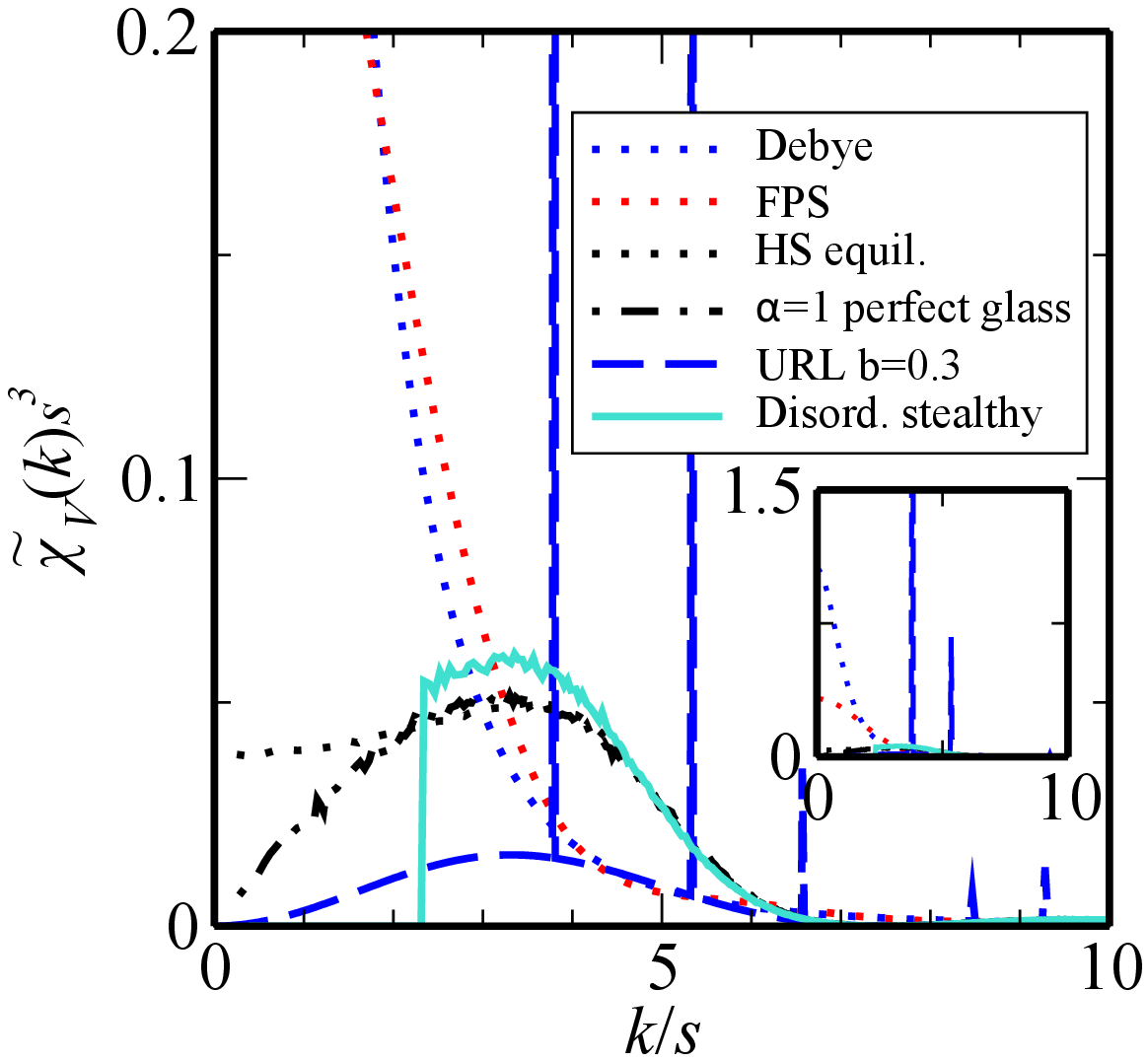}
}
 \caption{(a) Autocovariance functions $\chi_{_V}(r)$ for 3D models with $\phi_2=0.20$ and unit specific surface. (b) Spectral densities $\tilde{\chi}_{_V}(k)$ of the 3D models. The positions of the Bragg peaks for the cubic lattice packing agree exactly with those for the URL packing and are therefore not shown.}
  \label{chi_spec}
\end{figure*}

\subsection{Two-point statistics for the models}
To get a sense of how two-point statistics vary across the models considered here, we plot in Fig. \ref{chi_spec} the angular-averaged autocovariance functions $\chi_{_V}(r)$ and spectral densities $\tilde{\chi}_{_V}(k)$ for the 3D models, except for the BCC packings, for which the spectral density just consists of Bragg peaks. Figure \ref{chi_spec}(a) shows that all autocovariance functions have the same the slope at the origin due to our choice that each model possesses the same specific surface $s$. It is also obvious that all the sphere packings possess very similar forms of $\chi_{_V}(r)$ for $r\leq a=0.6/s$, because at low packing fractions, $S_2(\mathbf{r})$ at small $r$ is predominated by the probability that the two end points of $\mathbf{r}$ are in the same sphere. It has been shown that $s_2=d^2 \chi_{_V}(r)/dr^2|_{r=0}$ vanishes for unjammed sphere packings \cite{To02a}, which is the case for all the sphere packings in this study. On the other hand, it is clear from (\ref{debye}) and (\ref{chi_fps}) that $s_2>0$ for the Debye random media and FPS. We also note that all the hyperuniform media possess both positive and negative correlations, whereas the nonhyperuniform Debye random media and FPS possess only positive correlations.

Figure \ref{chi_spec}(b) shows the small-$k$ behaviors of the 3D models that we study. Among these models, the Debye random media has the largest value of $\lim_{k\rightarrow 0}\tilde{\chi}_{_V}(k)$, followed by FPS and equilibrium hard spheres. For the hyperuniform models, the perfect glass packing, the URL packing and the disordered stealthy sphere packing represent two-phase media with $\alpha=1, 2$ and $\infty$, respectively. As expected from Eq. (\ref{s_perturbedLatt}), $\tilde{\chi}_{_V}(k)$ for the URL packing is composed of a smooth function and Bragg peaks characteristic of the unperturbed cubic lattice packing. 

\subsection{Small-, intermediate- and large-time behaviors of the spreadabilities}

\begin{figure*}[!ht]
  \subfloat[]{
    \centering\includegraphics[trim=50 50 0 150, clip,width=5cm]{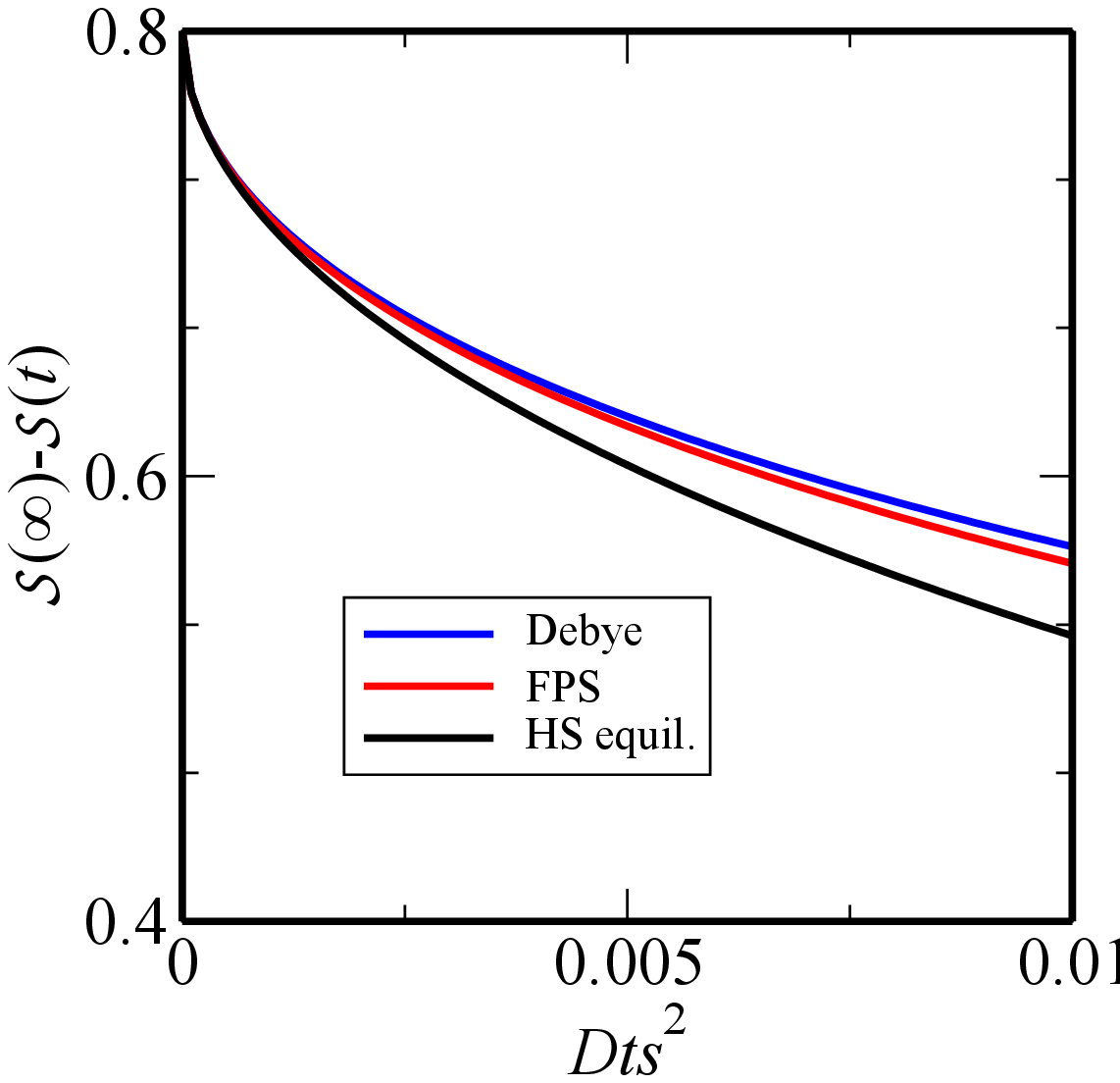}
}
\subfloat[]{
    \centering\includegraphics[trim=50 50 0 150, clip,width=5cm]{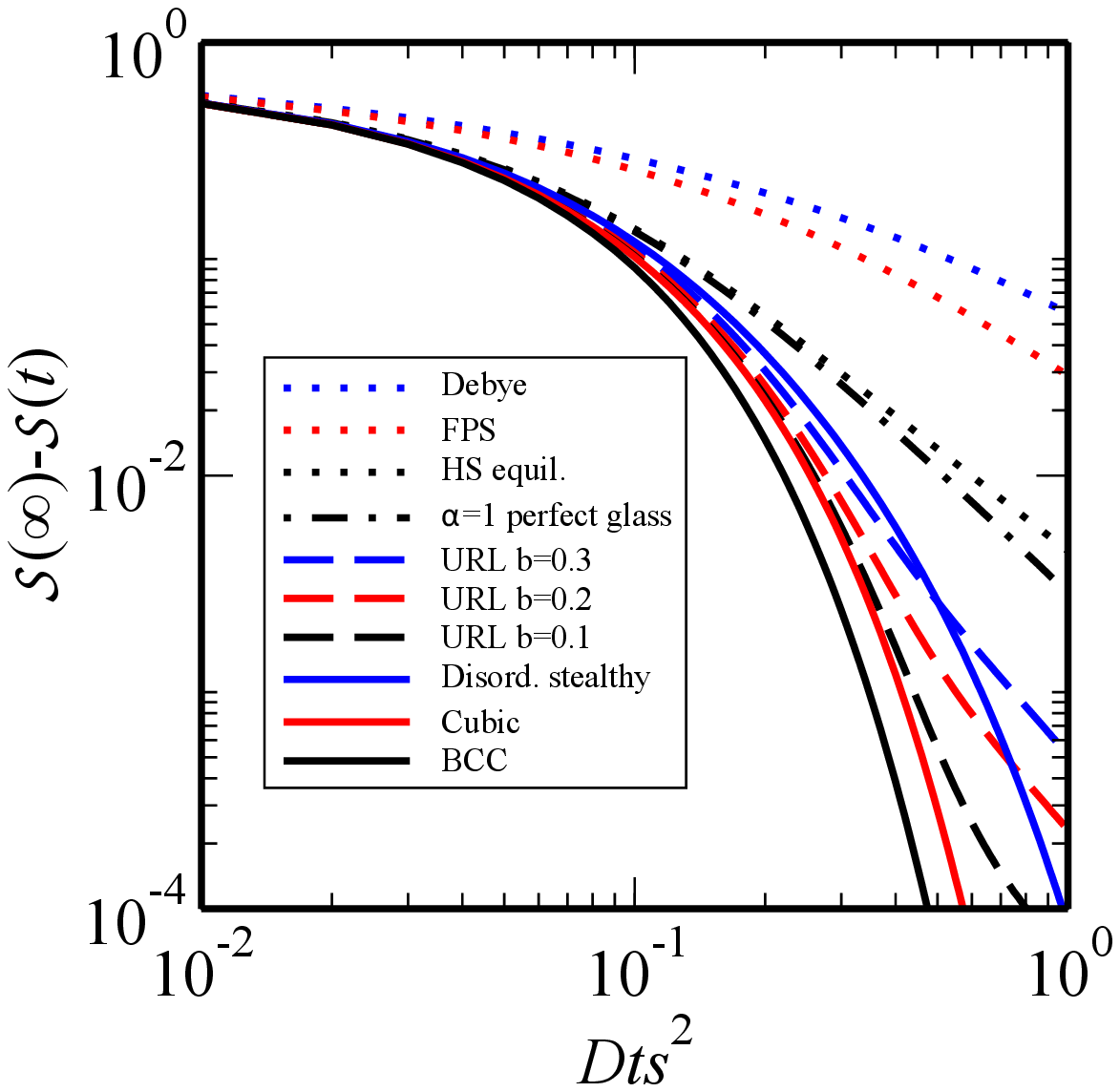}
}
\subfloat[]{
    \centering\includegraphics[trim=50 50 0 150, clip,width=5cm]{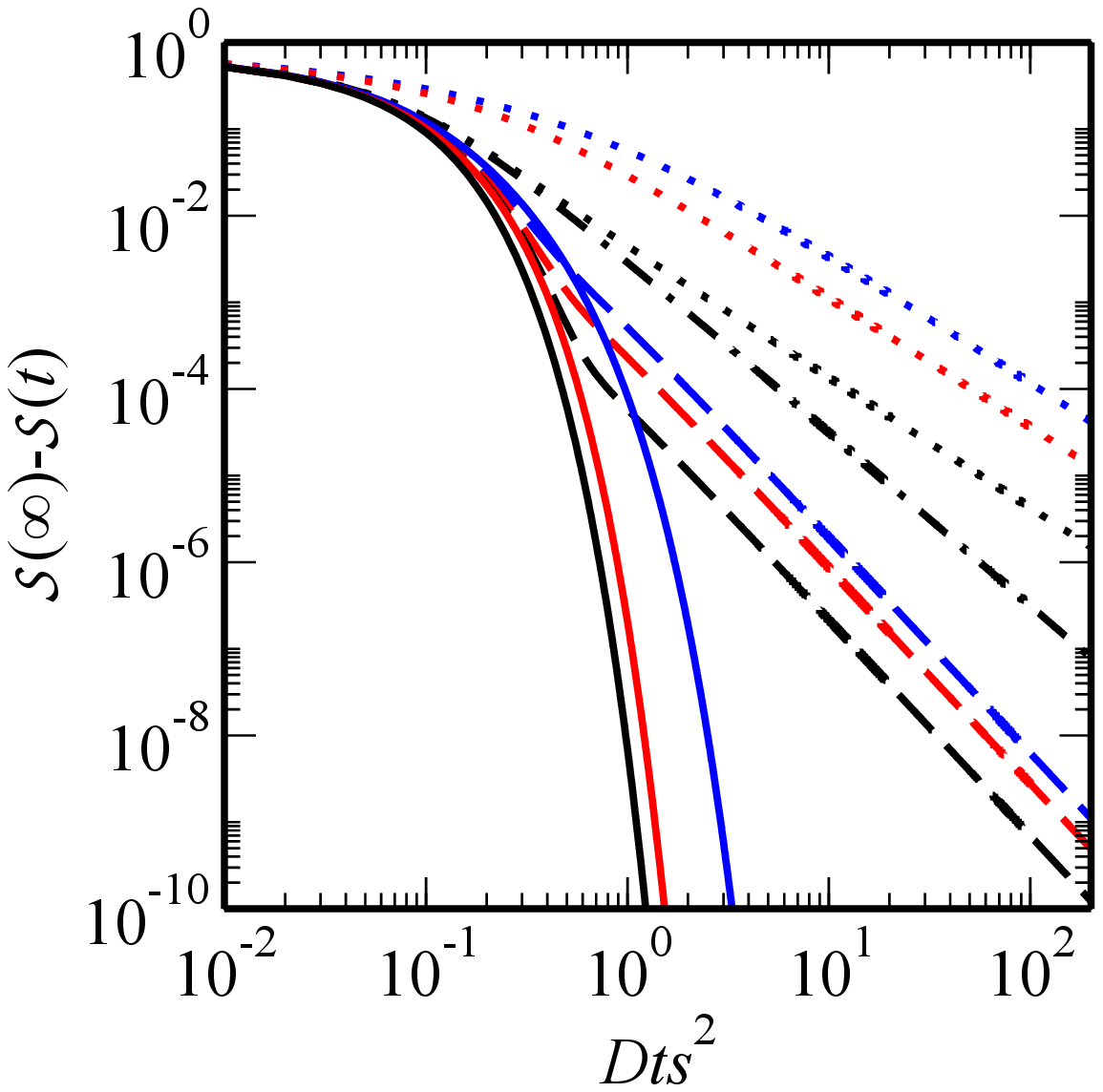}
}
  \caption{Spreadabilities for the 3D models considered with $\phi_2=0.20$ and unit specific surface. (a) Spreadabilities for the nonhyperuniform media at small dimensionless times ($0<Dts^2<0.01$).  (b) Spreadabilities for the 3D models at intermediate dimensionless times ($0.01<Dts^2<1$). (c) Spreadabilities for the 3D models at intermediate and large dimensionless times ($0.01<Dts^2<200$).}
  \label{3dspread}
\end{figure*}

Spreadabilities for the 3D models with $\phi_2=0.20$ are shown in Fig. \ref{3dspread}. Figure \ref{3dspread}(a) gives the spreadabilities associated with the three nonhyperuniform media in 3D at small dimensionless times ($0\leq Dts^2 \leq 0.01$).  Torquato \cite{To21d} derived short-time asymptotic expansion for $\mathcal{S}(t)$:
\begin{equation}
{\cal S}(t) = \frac{s}{\phi_2} \left(\frac{Dt}{\pi}\right)^{1/2} - \frac{2 \,d\,s_2}{\phi_2}\, (D t) + {\cal O}(Dt/a^2)^{3/2}.
\label{short-time}
\end{equation}
The initial decay rates are very similar for all models due to their identical specific surfaces. However, as soon as $Dts^2>0.002$, the second term in Eq. (\ref{short-time}) plays an important role in the distinguishing the spreadabilities. The positive values of $s_2$ for the Debye random media and FPS contribute to their slower spreadabilities compared to sphere packings, for which $s_2=0$.

Figure \ref{3dspread}(b) shows the spreadabilities for the 3D models at intermediate dimensionless times ($0.01\leq Dts^2\leq 1$). Table \ref{table_SAt1} lists the values of $\mathcal{S}(t)$ at dimensionless time $Dts^2=1$ for all the models. Note that the intermediate-time behavior of $\mathcal{S}(t)$ clearly captures intermediate-scale correlations of the two-phase media. Among the nonhyperuniform media, the intermediate-time spreadabilities are highly dependent on the coefficient $B=\tilde{\chi}_{_V}(0)$, which is positive and bounded \cite{To18a}. For example, 
Debye random media has a much larger value of $B$ than the other models, and its intermediate-time excess spreadability is also much larger. To explain this observation, we note that the spectral densities are nonnegative by definition and are continuous functions of $k$ for the nonhyperuniform models. Therefore, a larger value of $B$ implies that $\tilde{\chi}_{_V}(k)$ must be larger over a finite range $[0,k^*)$ of wavenumbers, where $k^*$ is some positive number. Integrating larger values of the integrand $\tilde{\chi}_{_V}(k)\exp[-k^2Dt]$ over the range $[0,k^*)$ in Eq. (\ref{spreadability_fourier}) then results in slower spreadabilities at intermediate and large times. We also find similar correspondence between $B$ and intermediate-time spreadabilities for the URL packings in which $\alpha=2$.

\begin{table}
\caption{Spreadability at $Dts^2=1$ for the 3D models with $\phi_2=0.20$.}
\begin{tabular}{ ||c|c|| } 
 \hline
 Model & $\mathcal{S}(\infty)-\mathcal{S}(1/Ds^2)$ \\ 
 \hline
 Debye & 0.0560  \\ 
 FPS & 0.0289  \\
 HS equil. & 0.00437  \\
 $\alpha=1$ perfect glass & 0.00285\\
 URL $b=0.3$ & $5.05\times 10^{-4}$  \\
 URL $b=0.2$ & $2.27\times 10^{-4}$  \\
 URL $b=0.1$ & $5.73\times 10^{-5}$  \\
 Disord. stealthy & $8.27\times 10^{-5}$\\
 Cubic & $1.78\times 10^{-7}$\\
 BCC & $5.72\times 10^{-9}$\\
 \hline
\end{tabular}
\label{table_SAt1}
\end{table}

\begin{center}
\begin{table}
\caption{Values of the coefficients $B$ and $C$ for the nonhyperuniform models and the URL packings in three dimensions with $\phi_2=0.20$.}
\begin{tabular}{ ||c|c|c|| } 
 \hline
 Model & $B$ & $C$ \\ 
 \hline
 Debye & 1.054 & 0.107 \\ 
 FPS & 0.3254 & 0.355 \\
 HS equil. & 0.0384 & 0.00389 \\
 URL $b=0.3$ & 0.00356 & $5.60\times 10^{-4}$ \\
 URL $b=0.2$ & 0.00158 & $2.49\times 10^{-4}$ \\
 URL $b=0.1$ & $3.95\times 10^{-4}$ & $6.25\times 10^{-5}$ \\
 \hline
\end{tabular}
\label{table_B}
\end{table}
\end{center}

Interestingly, Fig. \ref{3dspread}(b) shows that the spreadability curve for the disordered stealthy packing crosses those for the URL packings at intermediate dimensionless times. At small times, the latter spreadabilities are faster due to the presence of Bragg peaks in their associated spectral densities. On the other hand, at large times, the spreadability for the disordered stealthy packing is faster because the associated spectral density vanishes for a finite range of wavenumbers around the origin. The aforementioned crossing points represent the times beyond which the contribution from the stealthiness behavior of the disordered stealthy packing outcompetes the contribution from the Bragg peaks for the URL packings. The crossing time increases with decreasing $b$, since smaller $b$ values result in larger Bragg peaks for the URL packing [cf. (\ref{s_perturbedLatt})], and thus faster spreadabilities at small to intermediate times. Therefore, we have shown that in addition to probing microstructures on the very-small and very-large length scales, $\mathcal{S}(t)$ provides a powerful tool to study important intermediate-scale phenomena in heterogeneous media.

To highlight large-time behaviors of the spreadabilities, Fig. \ref{3dspread}(c) plots $\mathcal{S}(\infty)-\mathcal{S}(t)$ against $Dt$ on a log-log scale in the dimensionless time range $0.001<Dts^2<10$. It is clear that the large-$t$ behaviors of the spreadabilities is determined by the form of the spectral densities at small wavenumbers. Specifically, the excess spreadabilities for typical nonhyperuniform media (for which $\alpha=0$), which include Debye random media, fully penetrable spheres and equilibrium hard spheres, decay asymptotically as $t^{-3/2}$ \cite{To21d}. The spreadabilities for nonstealthy hyperuniform media (for which $0<\alpha<\infty$), including sphere packings derived from perfect glasses and URL, decay as $t^{-(3+\alpha)/2}$. The fastest decay rates are achieved by stealthy hyperuniform media (for which $\alpha=\infty$), including sphere packings derived from disordered stealthy point processes and Bravais lattices. The values of $\alpha$ can be very easily and accurately extracted from the slopes of the curves on the log-log plot at large times. 

For nonstealthy models with the same value of $\alpha$, Eq. (\ref{Spr_alpha}) predicts that the asymptotic decay rate of $\mathcal{S}(\infty)-\mathcal{S}(t)$ is directly proportional to the coefficient $B$. Table \ref{table_B} shows values of $B$ and $C$, where $C$ is the numerically determined coefficient for the long-time excess spreadability, i.e., $\mathcal{S}(\infty)-\mathcal{S}(t)\sim C/t^{(d+3)/2}$. The results confirm the predicted proportionality between $B$ and $C$. Among the three typical nonhyperuniform models ($\alpha = 0$), the Debye random media possesses the largest value of the coefficient $B=\tilde{\chi}_{_V}(0)$, and therefore has the slowest large-time spreadability among the models that we have considered. For the URL packings ($\alpha=2$), Table \ref{table_B} suggests that $B\propto C\propto b^2$. Indeed, it can be shown from (\ref{s_perturbedLatt}) that $\tilde{\chi}_{_V}(\mathbf{k})$ for URL packings is proportional to $b^2k^2$ at small $k$ \cite{Klatt2020}. For stealthy hyperuniform media, our calculations confirm the trend predicted in Ref. \cite{To21d} that the large-time spreadability is faster for models with larger values of $K$, where in the ordered case $K=Q_1$ is the first (smallest positive) Bragg wavenumber. Torquato \cite{To21d} showed that the long-time behavior of spreadabilities for both ordered and disordered stealthy media involve exponential decay rate $\exp(-K^2Dt)$, and hence possess faster decay rates than any hyperuniform medium governed by a power-law decay. Torquato \cite{To21d} has shown that the BCC lattice gives the optimal packing for the spreadability among all packings of identical spheres in three dimensions, as it possesses the largest value of $Q_1/s$. Our findings are consistent with
this optimal property of the BCC packing.

\subsection{Spreadabilities for URL packings}
\label{URL_spr}
Notably, we find that for the sphere packings derived from URL, the spreadabilities at small to intermediate dimensionless times are very similar to that of the unperturbed cubic-lattice packing, i.e. their initial decay rates are nearly exponential. However, at large $t$, the spreadabilities tend to a power-law decay, which is consistent with the small-wavenumber behavior ($\alpha=2$) of such packings. The nearly exponential decay up to intermediate times is due to the fact that the small- to intermediate-scale two-point correlations are mainly determined by the Bragg peaks in the URL packing \cite{Klatt2020}. On the other hand, it has been shown that the uncorrelated stochastic displacements of the lattice points degrade the stealthy hyperuniformity of the lattice into nonstealthy hyperuniformity described by the power-law (\ref{tildechi_alpha})
with $\alpha=2$ for URL \cite{Ki18a,Klatt2020}. We remark that for the URL packing with $b=0.1$, the snapshot in direct space [Fig. \ref{3dSnaps}(g)] is visually very similar to that of the unperturbed lattice packings [Fig. \ref{3dSnaps}(i)]. However, their spreadabilities are markedly different at large $t$. Therefore, $\mathcal{S}(t)$ sensitively captures long-range correlations that are not obvious in direct space.

\begin{figure*}[!ht]
\subfloat[]{
    \centering\includegraphics[trim=50 50 0 150, clip,width=8cm]{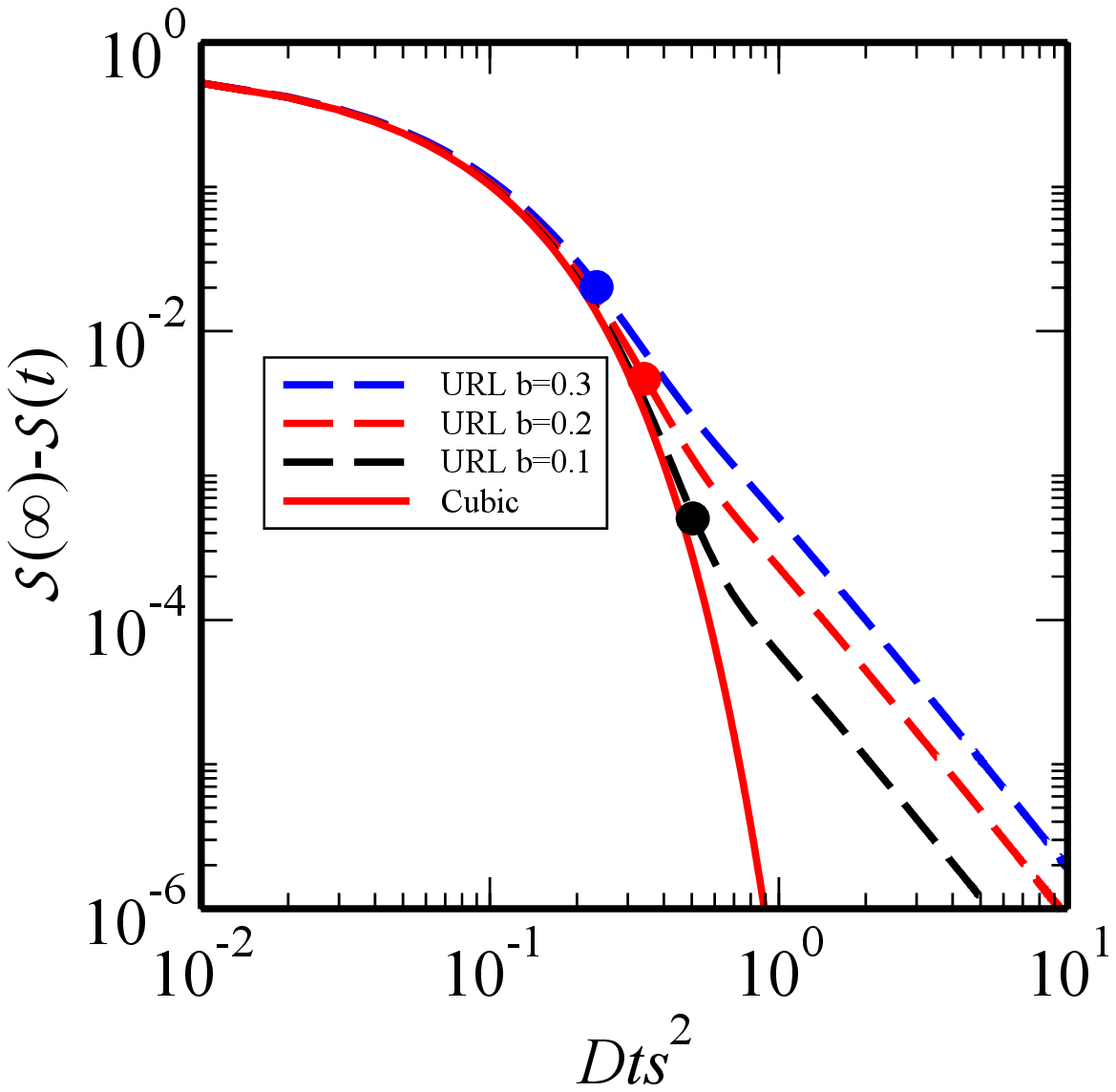}
}
\subfloat[]{
    \centering\includegraphics[trim=60 50 0 150, clip,width=8cm]{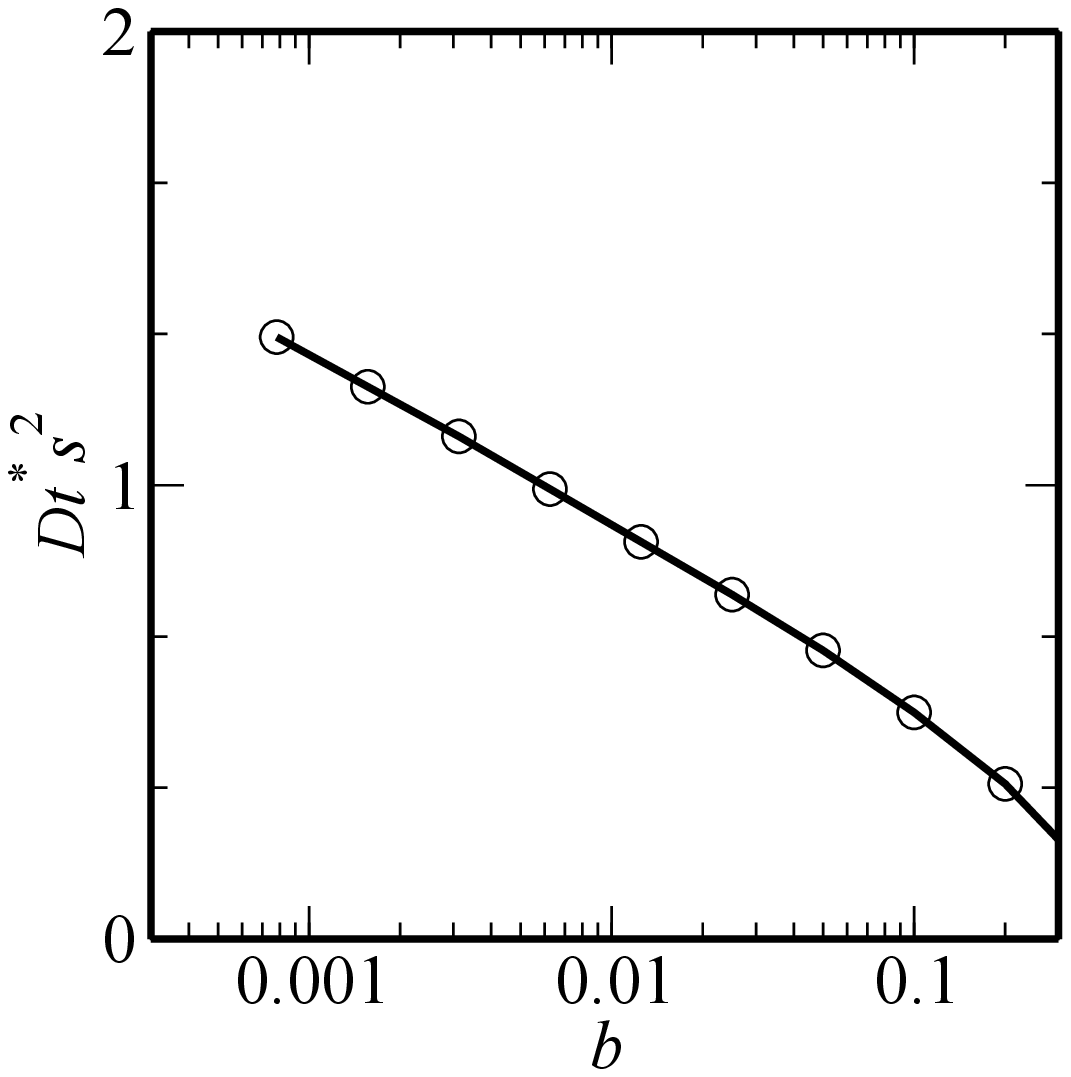}
}
  \caption{(a) Spreadabilities for 3D URL packings and the cubic-lattice packing with $\phi_2=0.20$ at intermediate dimensionless times ($0.01<Dts^2<10$). The transition time $t^*$ [see (\ref{t_transition})] for each value of $b$  are indicated by solid dots. (b) Plot of $t^*$ against $b$ for 3D URL packings with $\phi_2=0.20$.}
  \label{URL_transition_3D}
\end{figure*}

To quantitatively analyze the transition time $t^*$ of the spreadabilities for a URL packing, we note that for any monodisperse sphere packing derived from a perturbed lattice in which the perturbation characteristic function $\tilde{f}(\mathbf{k})$ for the lattice points are independent and identical, the excess spreadability can be obtained by substituting (\ref{chi-S}) into (\ref{spreadability_fourier}), where $S(k)$ in (\ref{chi-S}) is given by (\ref{s_perturbedLatt}), i.e.,
\begin{widetext}
\begin{equation}
\begin{split}
    &\mathcal{S}(\infty)-\mathcal{S}(t)=\\ &\phi_2\left[\int_0^{\infty} \tilde{\alpha}_2(k;a)(1 - |\tilde{f}(\mathbf{k})|^2)\exp(-k^2Dt) d \mathbf{k}+\sum_{\mathbf{Q}\ne \mathbf{0}}\frac{ \tilde{\alpha}_2(Q;a)}{v_1(a)}|\tilde{f}(Q)|^2\exp(-Q^2 Dt)\right].
\end{split}
    \label{spr_perturbedLatt}
\end{equation}
\end{widetext}

From relation (\ref{spr_perturbedLatt}), one can extract the transition time $t^*$ that separates small-time nearly exponential decay and large-time power-law decay of the spreadability by equating the Bragg-peak part and the diffuse part of the excess spreadability:
\begin{widetext}
\begin{equation}
    \int_0^{\infty} \tilde{\alpha}_2(k;a)(1 - |\tilde{f}(\mathbf{k})|^2)\exp(-k^2Dt^*) d \mathbf{k}=
    \sum_{\mathbf{Q}\ne \mathbf{0}}\frac{ \tilde{\alpha}_2(Q;a)}{v_1(a)}|\tilde{f}(Q)|^2\exp(-Q^2 Dt^*).
    \label{t_transition}
\end{equation} 
\end{widetext}
The specific form of $\tilde{f}(\mathbf{k})$ for URL packings derived from the hypercubic lattice for any space dimension $d$ and its angular average for $d=2,3$ is given in the \hyperref[sec:urlfTilde]{Appendix}. Figure \ref{URL_transition_3D} shows the spreadabilities for the URL packings with the corresponding transition times indicated by solid dots, as well as the spreadability for the unperturbed simple-cubic lattice packing. Note that on the log-log plots of $\mathcal{S}(t)$ for the URL packings, gradual changes of slope occur in the vicinity of $t^*$. Figure \ref{URL_transition_3D}(b) plots the values of $t^*$ against $b$ for 3D URL packings with $\phi_2=0.20$, in which we observe that $t^*$ is a decreasing function of $b$ and has a logarithmic divergence in the limit $b\rightarrow 0$, i.e., $t^*\sim -\ln b$. To derive this logarithmic divergence behavior, we first presume that $t^*$ is large in the limit $b\rightarrow 0$. The factors multiplying $\exp(-k^2Dt^*)$ on the left-hand side of (\ref{t_transition}) can then be replaced by their small-$k$ behaviors, i.e., $1-|\tilde{f}(\mathbf{k})|^2\rightarrow b^2k^2$  \cite{Klatt2020}, and $\tilde{\alpha}_2(k;a)\rightarrow \exp(-ck^2)$, where $c$ is a constant \cite{To18a}. Therefore, the left-hand side of Eq. (\ref{t_transition}) has the small-$b$ limit
\begin{equation}
\begin{split}
   \int_0^\infty \exp(-ck^2)b^2k^2\exp(-k^2Dt^*)d\mathbf{k}\\=\frac{b^2\exp(-ck^2)\sqrt{\pi}}{4t^{3/2}}.
    \label{transition_lhs}
\end{split}
\end{equation}
The small-$b$ behavior of the sum on right-hand side of (\ref{t_transition}) is dominated by the Bragg peak with the smallest wavenumber $Q_1$. Therefore, the right-hand side of (\ref{t_transition}) has the small-$b$ limit
\begin{equation}
    \frac{\tilde{\alpha}_2(Q_1;a)Z(Q_1)}{v_1(a)}\exp(-Q_1^2Dt^*),
    \label{transiton_rhs}
\end{equation}
where $Z(Q_1)$ is the coordination number at radial distance $Q_1$. Equating (\ref{transition_lhs}) and (\ref{transiton_rhs}) yields
\begin{equation}
    2\ln b \sim \frac{3\ln t^*}{2}-Q_1^2Dt^*,
    \label{t*_small_b}
\end{equation}
which corresponds to a logarithmic divergence of $t^*$ as $b\rightarrow 0$. Our numerical solution of Eq. (\ref{t_transition}) at small $b$ confirms the asymptotic relation (\ref{t*_small_b}).

\subsection{Effects of dimensionality on the spreadability}
\begin{figure*}[!ht]
\subfloat[]{
    \centering\includegraphics[trim=0 50 0 100, clip,width=5cm]{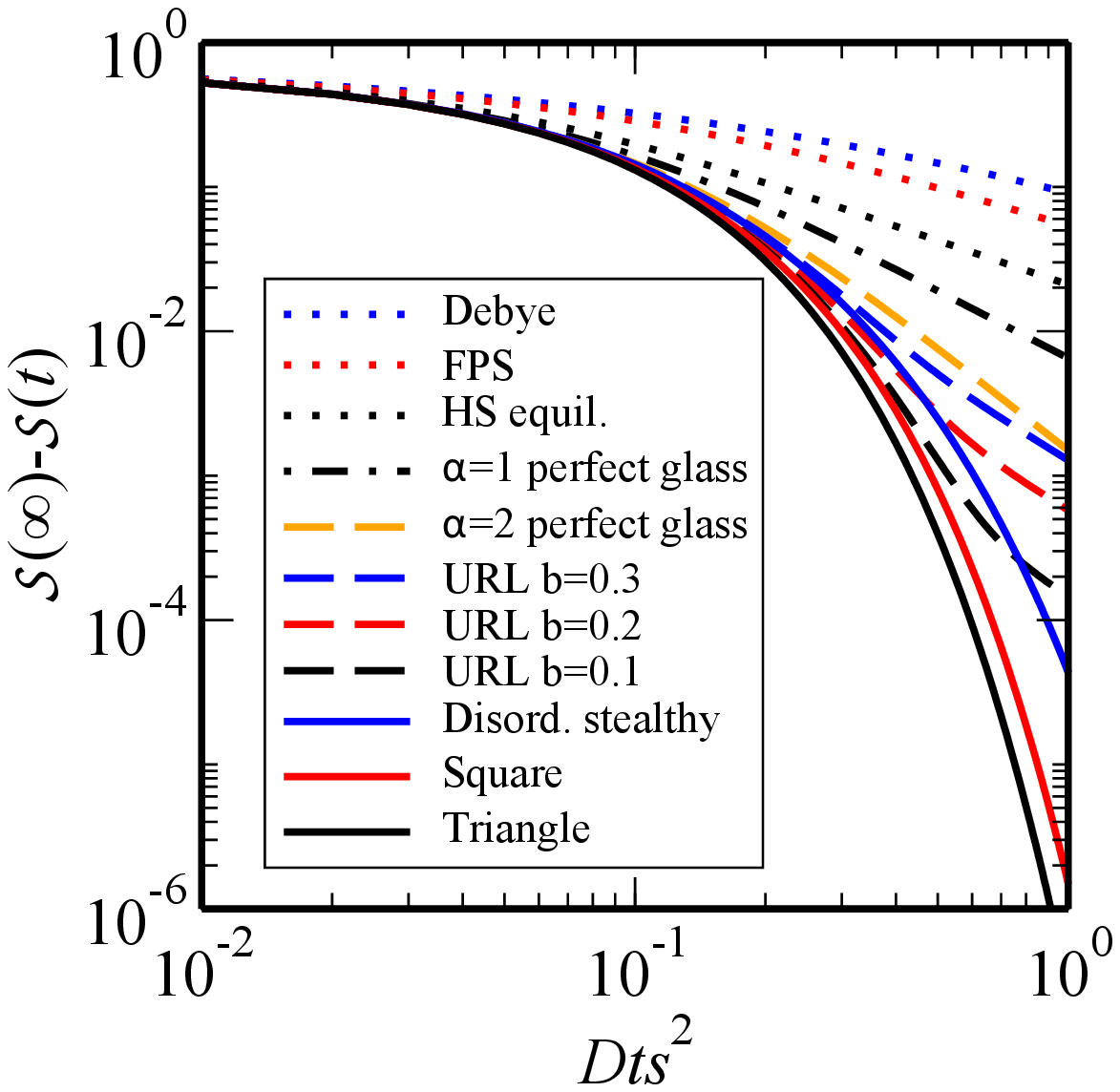}
}
\subfloat[]{
    \centering\includegraphics[trim=0 50 0 100, clip,width=5cm]{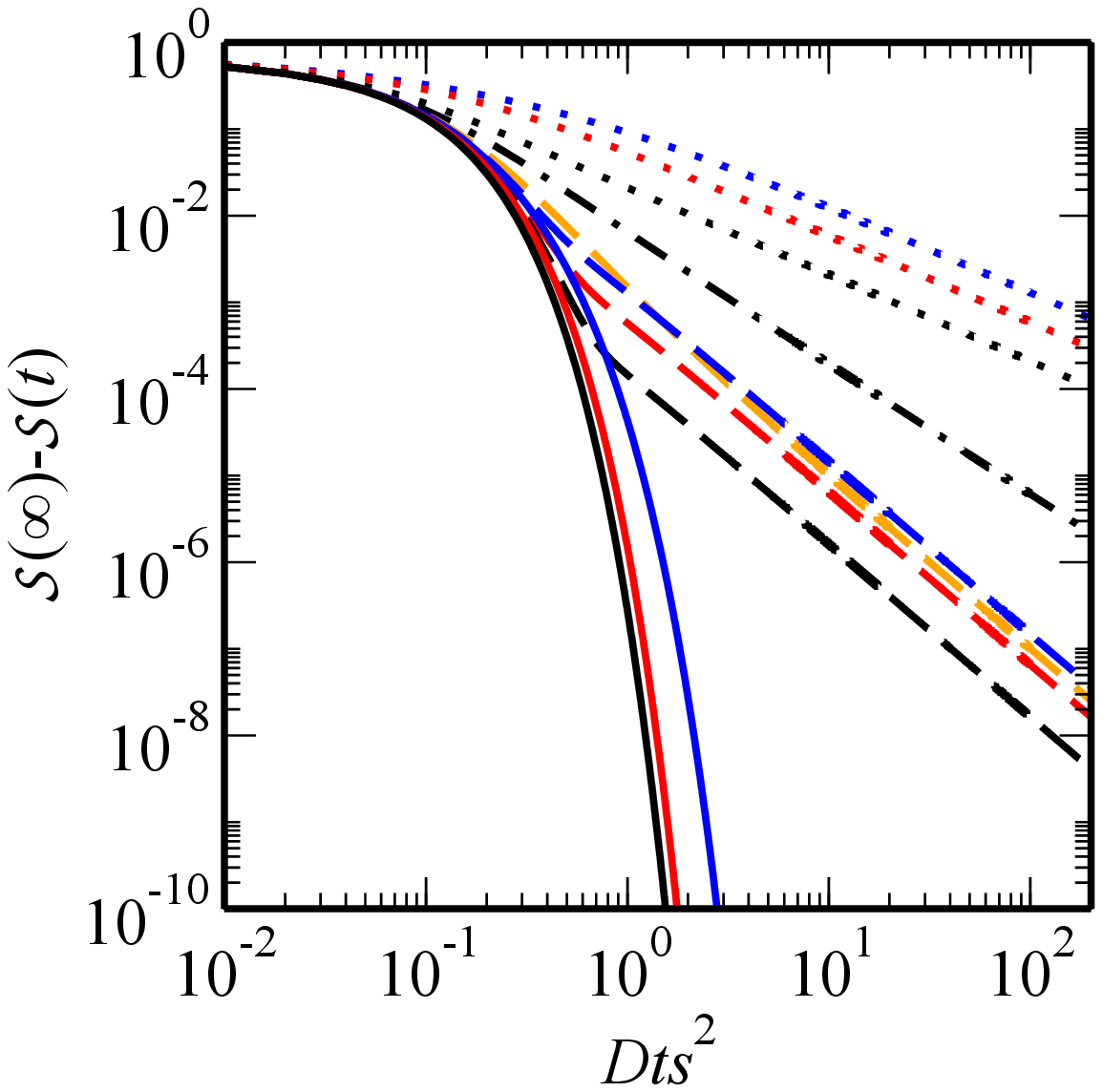}
}
\subfloat[]{
    \centering\includegraphics[trim=0 50 0 100, clip,width=5cm]{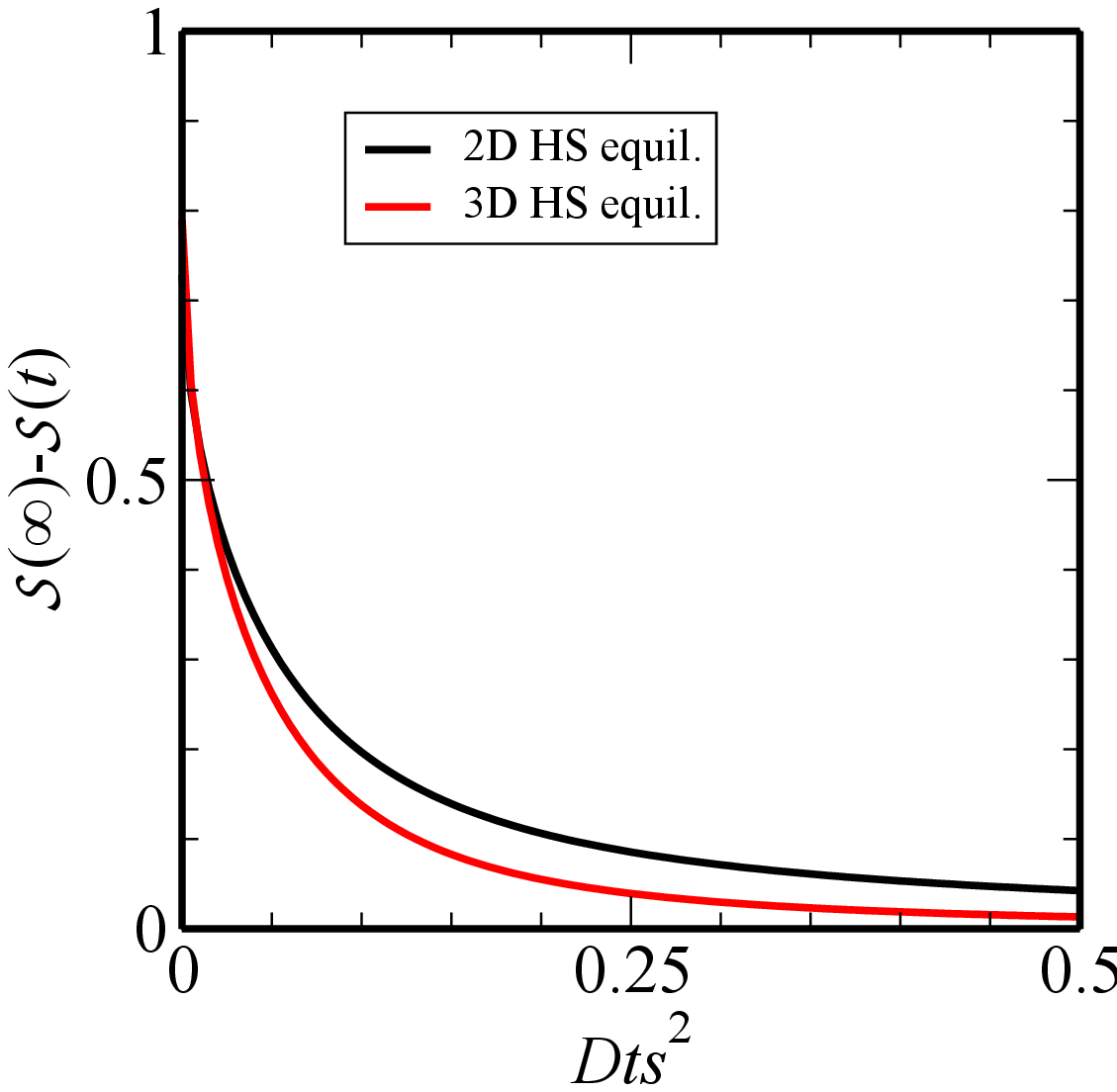}
}
  \caption{(a) Spreadabilities of two-phase media in two dimensions with $\phi_2=0.25$ and unit specific surface at intermediate dimensionless times ($0.01<Dts^2<1$). (b) Spreadabilities of two-phase media in two dimensions with $\phi_2=0.25$ and unit specific surface at intermediate and large dimensionless times ($0.01<Dts^2<200$). (c) Spreadabilities for equilibrium hard spheres for $d=2, \phi_2=0.25$ and $d=3,\phi_2=0.20$ with unit specific surface at small and intermediate dimensionless times ($0<Dts^2<0.5$).}
  \label{2dspread}
\end{figure*}

To compare spreadabilities across dimensions, we show spreadabilities of the 2D models in Fig. \ref{2dspread}. We observe in Fig. \ref{2dspread}(a) that for the same value of $\alpha$, the excess spreadability decays asymptotically as $t^{-(2+|\alpha|)/2}$, slower than that for 3D models. This confirms the large-$t$ trend predicted in Ref. \cite{To21d}, i.e., $\mathcal{S}(\infty)-\mathcal{S}(t)\sim t^{-(d+\alpha)/2}$. Again, we notice that the excess spreadabilities for the URL packings transform from exponential decay at small and intermediate $t$ to power-law decay at large $t$. Figure \ref{2dspread}(b) compares the spreadabilities of equilibrium hard spheres at small to intermediate $t$. The small-time decay of the excess spreadability is faster for the 3D model. This is evident from Eq. (\ref{short-time}), as $s/\phi_2$ is larger for the 3D model. 

\section{Algorithm for extracting long-time behaviors from Time-Dependent Diffusion Measurements}
\label{algo}
Based on our study of spreadabilities of the idealized models, we introduce here an algorithm that extracts large-scale behavior from time dependent diffusion behaviors as measured by spreadability, or equivalently, by NMR or MRI data. For simplicity, we first describe the algorithm using the spreadability. Subsequently, we discuss the adaption of the algorithm to NMR and MRI measurements using the precise mappings identified in Ref. \cite{To21d}. For purposes of illustration, we treat the spreadability data as if they were directly given by diffusion measurements when applying the algorithm, despite the fact that we computed $\mathcal{S}(t)$ from two-point correlations in this study.

To extract the asymptotic form of any spreadability data $\mathcal{S}(t)$, we introduce a ``set-in'' time $t_S(\epsilon)$, defined as the time such that $|\mathcal{S}(t)-\mathcal{S}_l(t)|=\epsilon$ for $t=t_S$ and $|\mathcal{S}(t)-\mathcal{S}_l(t)|<\epsilon$ for all $t>t_S$. Here, $\mathcal{S}_l(t)$ is the long-time excess spreadability given in Ref. \cite{To21d}:
\begin{widetext}
\begin{equation}
    \mathcal{S}(\infty)-\mathcal{S}_l(t)=Ct^{-\varphi} \qquad \text{Nonstealthy media},
    \label{nonstealthy_longtime}
\end{equation}
\begin{equation}
    \mathcal{S}(\infty)-\mathcal{S}_l(t)=C\exp(-\varphi t)/t \qquad \text{Disordered stealthy media},
    \label{disordered_stealthy_longtime}
\end{equation}
\begin{equation}
    \mathcal{S}(\infty)-\mathcal{S}_l(t)=C\exp(-\varphi t) \qquad \text{Ordered media},
    \label{ordered_longtime}
\end{equation}
\end{widetext}
where $\varphi$ and $C$ are parameters to be determined from the spreadability data. The self-consistent relations (\ref{nonstealthy_longtime}), (\ref{disordered_stealthy_longtime}) and (\ref{ordered_longtime}) can be solved via the following iterative procedure. 

\begin{enumerate}
    \item Set $i=0$ and choose initial guesses of  $t_{S,i}, \varphi_i$ and $C_i$ as well as the stopping criterion $\sigma$.
    \item Find $\varphi_{i+1}$, $C_{i+1}$ by fitting the spreadability for $t>t_{S,i}$ and obtain the large-time spreadability approximant $S_{l,i}(t)$ using (\ref{nonstealthy_longtime}), (\ref{disordered_stealthy_longtime}) or (\ref{ordered_longtime}). 
    \label{fit}
    \item Find $t_{S,i+1}$, which is the time such that $|\mathcal{S}(t_{S,i+1})-\mathcal{S}_{l,i+1}(t_{S,i+1})|=\epsilon$ and $|\mathcal{S}(t)-\mathcal{S}_{l,i}(t)|<\epsilon$ for all $t>t_{S,i+1}$.
    \item If $|\varphi_{i+1} - \varphi_{i}|<\sigma$, stop and set $t_S=t_{S,i+1}, \varphi = \varphi_{i+1}, C=C_{i+1}$. If not, set $i=i+1$ and go to Step~\ref{fit}.
\end{enumerate}

Because it is not known \textit{a priori} whether a two-phase media is nonstealthy, disordered stealthy or ordered, the aforementioned procedure must be executed three different times, using Eq. (\ref{nonstealthy_longtime}), (\ref{disordered_stealthy_longtime}) and (\ref{ordered_longtime}) alternatively each time. The equation that yields the smallest $t_S$ is then accepted as the functional form that best describes the asymptotic behavior of the spreadability. Specifically, assuming nonstealthy media, $\varphi_{i+1}$ and $C_{i+1}$ in Step \ref{fit} are obtained from a linear regression on the plot of $\ln[\mathcal{S}(\infty)-\mathcal{S}(t)]$ against $\ln(t)$. Assuming disordered stealthy media, the linear regression is performed on the plot of $t\ln[\mathcal{S}(\infty)-\mathcal{S}(t)]$ against $t$. Assuming ordered media, the linear regression is performed on the plot of $\ln[\mathcal{S}(\infty)-\mathcal{S}(t)]$ against $t$. We chose the stopping criterion to be $\sigma = 10^{-6}$ for all executions of the algorithm.

Table \ref{set_in_2D} and \ref{set_in_3D} list the values of $t_S(10^{-6})$, $\varphi$ and $C$ determined from the algorithm for the 2D and 3D models studied in this work, as well as the expected values of $\varphi$ and $C$. These parameters are known exactly for Debye random media, FPS, URL packings and Bravais-lattices packings, whereas those for the other models in Sec. \ref{models} are inferred from the corresponding spectral densities. As these idealized models span a wide range of nonhyperuniform and hyperuniform classes, they serve as good benchmarks for testing the general capability of the algorithm to accurately extract $\varphi$ and $C$ from diffusion data of realistic heterogeneous media. Table \ref{set_in_2D} and \ref{set_in_3D} show that for all the models that we study, the values of $\varphi$ and $C$ extracted using our proposed algorithm agree closely with their expected values. In cases in which these parameters are exactly known, the extracted $\varphi$ values deviate less than 1\% from the exact ones and the extracted $C$ values deviate less than 5\% from the exact ones. Indeed, we find that the extracted $\varphi$ and $C$ converge to their exact values as $\epsilon\rightarrow 0$, and that $\varphi$ can be estimated with reasonable accuracy (less than $10\%$ error) as long as $\epsilon < 10^{-4}$.

We also observe a general trend that $t_S$ decreases with increasing nonhyperuniform or hyperuniform class. Among models with the same value of $\alpha$, $t_S$ decreases with decreasing $B$. Among the stealthy models, $t_S$ decreases with increasing $K$. This is because for models with a higher degree of hyperuniformity, the corresponding spectral densities can usually be approximated by the power law (\ref{tildechi_alpha}) over a larger range of wavevectors. However, we point out that for URL packings with extremely small $b$, the ``set-in'' time of the power-law decay is expected to increase due to large Bragg peaks.

We remark that due to the connection between the spreadability and NMR measurements, including the pulsed field gradient spin-echo (PFGSE) amplitude  $\mathcal{M}(k,t)$ \cite{Mit92b, Se94} and the MRI-measured water diffusion in biological media \cite{No14}, our algorithm can be directly applied on NMR or MRI data to infer large-scale microstructural information of real materials. Specifically, consider a fluid-saturated porous medium, which invariably contains paramagnetic impurities at the interface, resulting generally partially absorbing boundary conditions. Torquato \cite{To21d} found that one can map the spreadability problem to the PFGSE amplitude problem via the transformations $\mathcal{S}(\infty)-\mathcal{S}(t)\rightarrow\mathcal{M}(\mathbf{q}=0,t)-\phi_2$ and $D\rightarrow\mathcal{D}(t)$, where $\phi_2$ is the porosity and $\mathcal{D}(t)$ is the time-dependent diffusion coefficient for the porous medium. For diffusion-weighted MRI of water-saturated biological media, one has a similar mapping $\mathcal{S}(\infty)\rightarrow \mathcal{D}_e=\mathcal{D}(\infty)$ and $\mathcal{S}(t)\rightarrow\mathcal{D}(t)$ \cite{To21d}. Therefore, our algorithm can be directly adapted to analyze experimental data for PFGSE or MRI to extract the coefficients $\varphi$ and $C$.

\begin{table*}[htbp]
\centering
\caption{``Set-in'' times $t_S$ and large-time parameters $\varphi$ and $C$ for the  spreadabilities of the 2D models considered in this work.}
\begin{minipage}{\textwidth}
\renewcommand*{\thempfootnote}{\fnsymbol{mpfootnote}}
\begin{tabular}{ ||c|c|c|c|c|c|c|| } 
 \hline
Model & $\phi_2$ & $Dt_S(10^{-6})s^2$ & $\varphi$ & $C$ & Expected $\varphi$ & Expected $C$
 \\ 
 \hline
 Debye & 0.25 & 309 & 1.003 & 0.133 & 1 & 0.130 \\ 
 FPS & 0.25 & 43.1 & 0.999 & 0.0592 & 1 & 0.0594 \\
 HS equil. & 0.25 & 17.2 & 1.002 & 0.0211 & 1 & 0.0209 \cite{simu} \\
 Perfect glass $\alpha=1$ & 0.25 & 13.06 & 1.499 & $6.12\times 10^{-3}$ & 3/2 & $6.15\times 10^{-3}$ \cite{simu} \\
 Perfect glass $\alpha=2$ & 0.25 & 5.67 & 2.013 & $1.06\times 10^{-3}$ & 2 & $9.87\times 10^{-4}$ \cite{simu}] \\
 URL $b=0.3$ & 0.25 & 4.71 & 1.997 & $1.44\times 10^{-3}$ & 2 & $1.48\times 10^{-3}$ \\
 URL $b=0.2$ & 0.25 & 3.63 & 1.997 & $6.41\times 10^{-4}$ & 2 & $6.56\times 10^{-4}$ \\
 URL $b=0.1$ & 0.25 & 2.34 & 1.996 & $1.60\times 10^{-4}$ & 2 & $1.64\times 10^{-4}$ \\
 Disord. stealthy & 0.25 & 1.04 & 6.662 & 0.0329 & 6.648 \cite{simu} & 0.0370 \cite{simu} \\
 Square & 0.25 & 0.710 & 12.55 & 0.418 & 12.57 & 0.430\\
 Triangle & 0.25 & 0.600 & 14.450 & 0.548 & 14.50 & 0.558\\
 \hline
\end{tabular}
\label{set_in_2D}
\end{minipage}
\end{table*}

\begin{table*}[htbp]
\centering
\caption{``Set-in'' times $t_S$ and large-time parameters $\varphi$ and $C$ for the  spreadabilities of the 3D models considered in this work.}
\begin{minipage}{\textwidth}
\renewcommand*{\thempfootnote}{\fnsymbol{mpfootnote}}
\begin{tabular}{ ||c|c|c|c|c|c|c|| } 
 \hline
Model & $\phi_2$ & $Dt_S(10^{-6})s^2$ & $\varphi$ & $C$ & Expected $\varphi$ & Expected $C$
 \\ 
 \hline
 Debye & 0.2 & 110 & 1.494 & 0.113 & 3/2 & 0.118 \\ 
 FPS & 0.2 & 28.1 & 1.499 & 0.0361 & 3/2 & 0.0365 \\
 HS equil. & 0.2 & 7.23 & 1.500 & $4.30\times 10^{-3}$ & 3/2 & $4.31\times 10^{-3}$ \cite{simu}\\
 Perfect glass $\alpha=1$ & 0.2 & 4.02 & 2.001 & $3.14\times 10^{-3}$ & 2 & $3.14\times 10^{-4}$ \cite{simu} \\
 URL $b=0.3$ & 0.2 & 3.39 & 2.495 & $6.06\times 10^{-4}$ & 5/2 & $6.25\times 10^{-4}$ \\
 URL $b=0.2$ & 0.2 & 2.66 & 2.494 & $2.68\times 10^{-4}$ & 5/2 & $2.78\times 10^{-4}$ \\
 URL $b=0.1$ & 0.2 & 1.80 & 2.494 & $6.69\times 10^{-5}$ & 5/2 & $6.94\times 10^{-5}$ \\
 Disord. stealthy & 0.2 & 1.06 & 5.377 & 0.0176 & 5.341 \cite{simu} & 0.0168 \cite{simu} \\
 Cubic & 0.2 & 0.410 & 14.43 & 0.388 & 14.43 & 0.387\\
 BCC & 0.2 & 0.380 & 18.18 & 0.554 & 18.18 & 0.556\\
 \hline
 Debye & 0.157 & 48.1 & 1.496 & 0.0686 & 3/2 & 0.0704 \\
 Fontainebleau & 0.157 & 24.9 & 1.501 & 0.0379 & - & - \\
 FPS & 0.157 & 21.9 & 1.499 & 0.0253 & 3/2 & 0.0255 \\
 \hline
\end{tabular}
\label{set_in_3D}
\end{minipage}
\end{table*}

\section{Application to Fontainebleau sandstone}
\label{sec:fontainebleau}
In this section, we apply the lessons learned from the spreadabilities of the idealized models to analyze the structural characteristics of a sample Fontainebleau sandstone \cite{Co96}. Because Fontainebleau sandstone is a porous medium with a pure mineral composition (almost 100\% quartz) and free of any clay, it has frequently been used to study the relation between pore geometry and effective properties. \cite{Co96,Bied2002,Chen2016,Ho16,Saadi2017,Sun2018}. Understanding its microstructures is of importance in many applications, including benchmark flow experiments \cite{Saadi2017,Ho16}, petroleum engineering \cite{Bied2002,Chen2016} and geological dating \cite{Mi17}. The morphology and physical properties (e.g. fluid permeability) of Fontainebleau sandstone samples have been accurately studied via X-ray tomographic imaging \cite{Co96,Sun2018}. Here, we further characterize their microstructures via diffusion spreadability measurements using the analysis described in Sec. \ref{res} and \ref{algo}. 

\begin{figure*}[!ht]
\subfloat[]{
    \centering\includegraphics[trim=0 0 0 -20, clip, width=5cm]{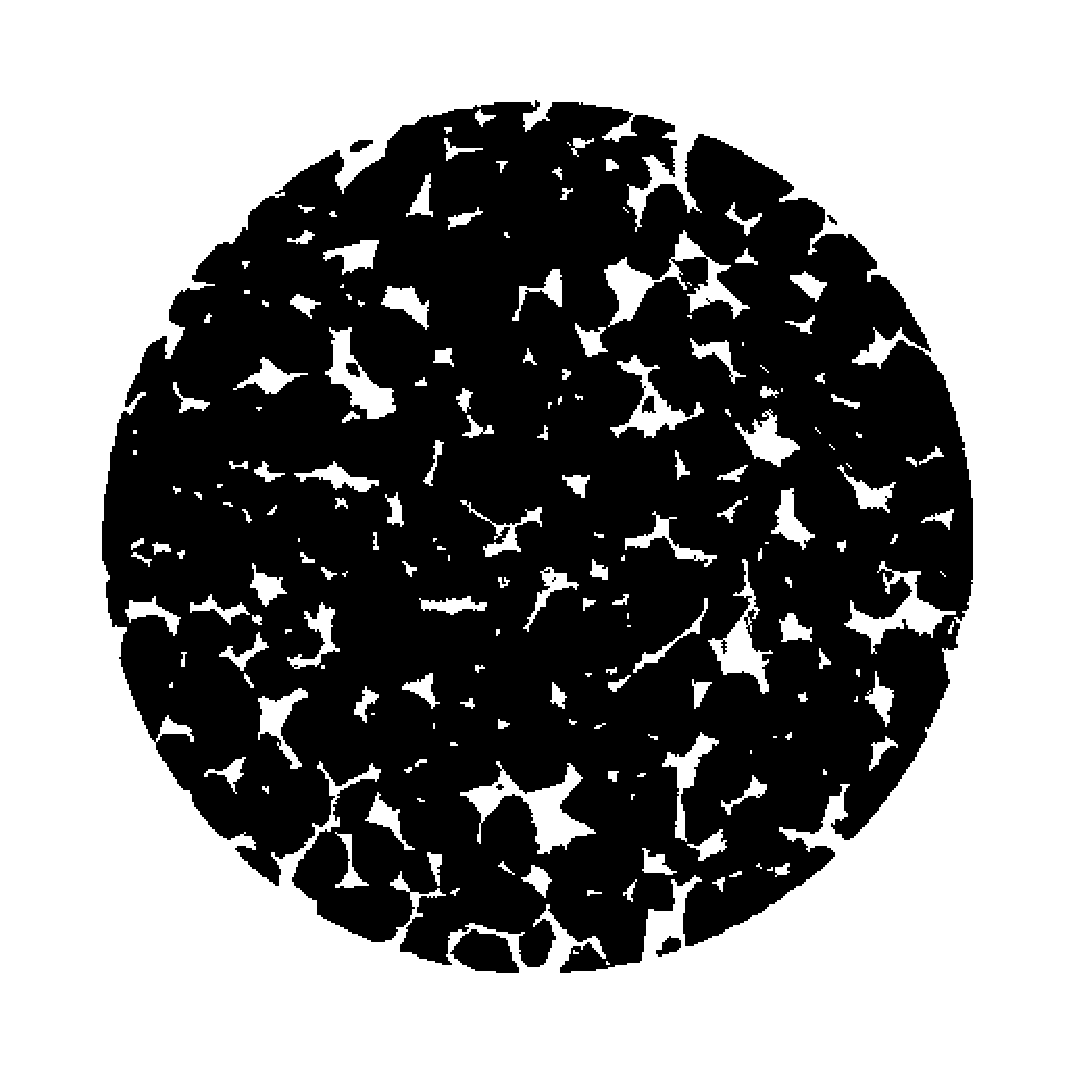}
}
\subfloat[]{
    \centering\includegraphics[trim=50 50 0 100, clip, width=5cm]{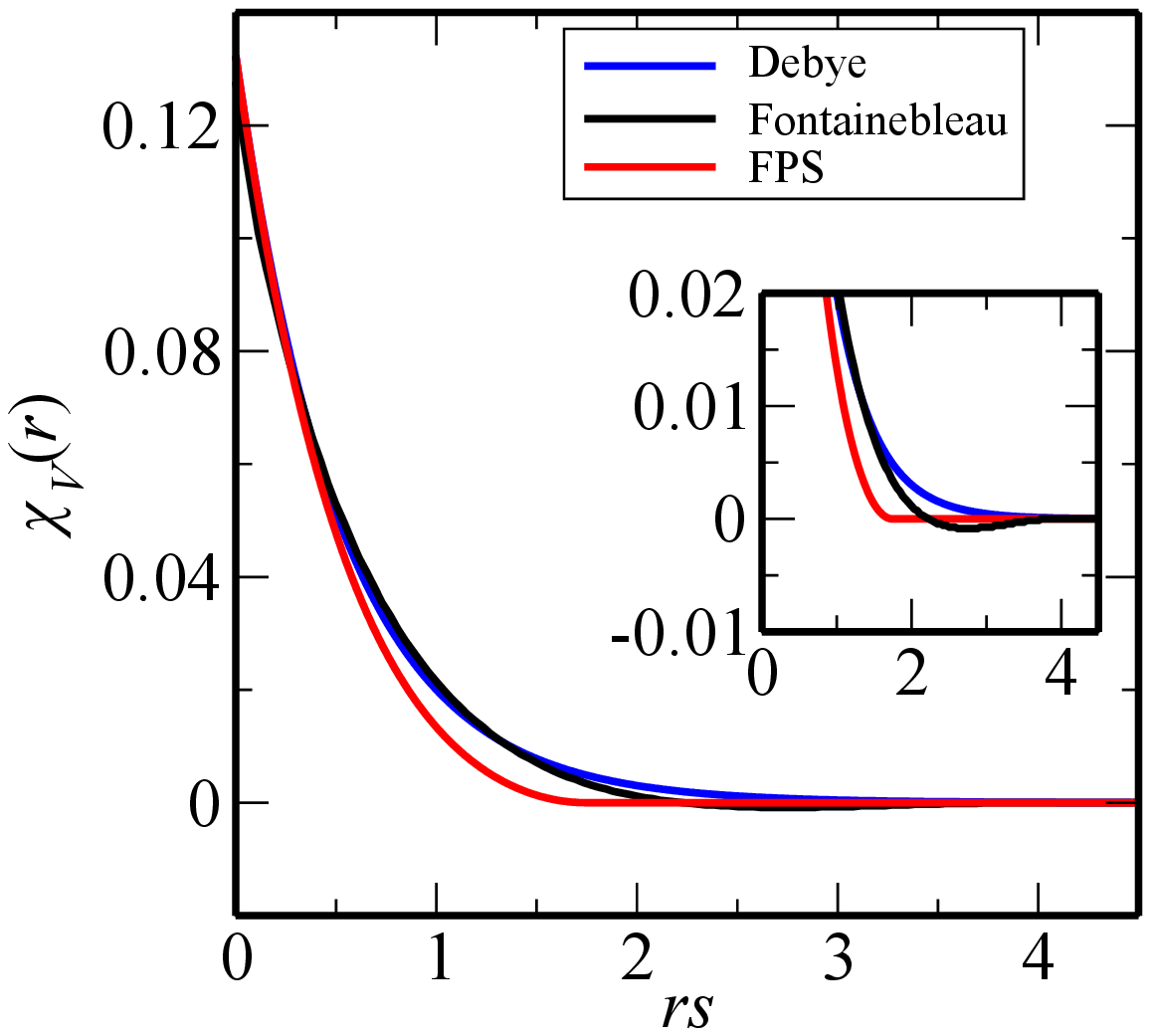}
}
\subfloat[]{
    \centering\includegraphics[trim=50 50 0 100, clip, width=5cm]{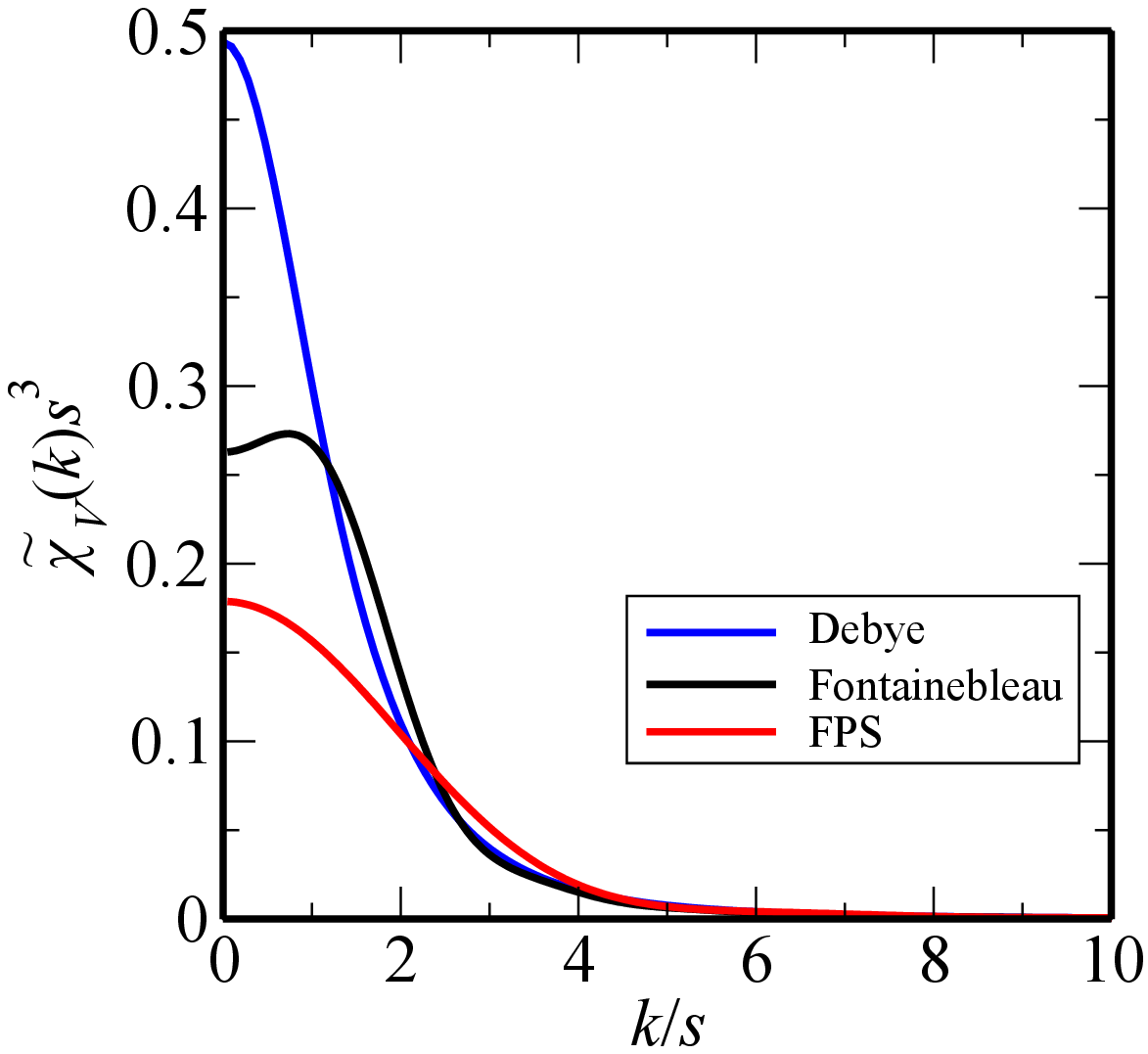}
}
  \caption{(a) Sample filtered slice of Fontainebleau sandstone. The black region corresponds to the grain phase. The diameter of the cylindrical core is 3 mm with a voxel resolution of 7.5 $\mu$m. This figures is reproduced from the one in Ref. \cite{Co96}. (b) Autocovariance functions for Fontainebleau sandstone \cite{Co96}, Debye random media and fully penetrable spheres with $\phi_2=0.157, s=0.0154$ $\mu$m$^{-1}$. (c) Spectral densities for Fontainebleau sandstone, Debye random media and fully penetrable spheres.}
  \label{fontainebleau_chi}
\end{figure*}

Figure \ref{fontainebleau_chi}(a) shows a sample filtered slice of Fontainebleau sandstone \cite{Co96}. Figure \ref{fontainebleau_chi}(b) and (c) show the autocovariance function and spectral density, respectively, for a sandstone sample with porosity $\phi_2=0.157$ and specific surface $s=0.0154\mu\text{m}^{-1}$, which we obtained from the corresponding two-point correlation function $S_2(r)$ published in Ref. \cite{Co96}. We observe that $\chi_{_V}(r)$ for Fontainebleau sandstone contains slightly negative correlations in the range $2.2<rs<3.7$, and vanishes identically for $rs\geq 3.7$, which indicates that the sample
is not hyperuniform. The negative correlations in $\chi_{_V}(r)$ corresponds to a maximum away from the origin in $\tilde{\chi}_{_V}(k)$. Indeed, the plot of the spectral density, shown in Fig. \ref{fontainebleau_chi}(c), definitively enables us to conclude that the sandstone is nonhyperuniform of the typical kind; see Eq. (\ref{sigma-nonhyper}). Therefore, it is instructive to compare its two-point correlations to those for the idealized typical nonhyperuniform models with $\phi_2=0.157$, including the Debye random media and FPS. Figure \ref{fontainebleau_chi}(b) shows that $\chi_{_V}(r)$ for the sandstone closely resemble that for the Debye random media at small to intermediate values of $rs$. Figure \ref{fontainebleau_chi}(c) shows that the value of $B = \tilde{\chi}_{_V}(0)$ for the sandstone lies in between those for the Debye random media and FPS. As in the case of FPS, the fact that $\chi_{_V}(r)$ for the sandstone has finite support enables us to accurately compute the corresponding spreadability using Eq. (\ref{spreadability_direct}) in this study. As we have stressed earlier, we can infer structural characteristics across length scales by direct dynamic measurements via the time-dependent spreadability or NMR/MRI techniques \cite{Mit92b,Se94,No14}.

\begin{figure}[!ht]
    \centering\includegraphics[trim=50 50 0 100, clip, width=6cm]{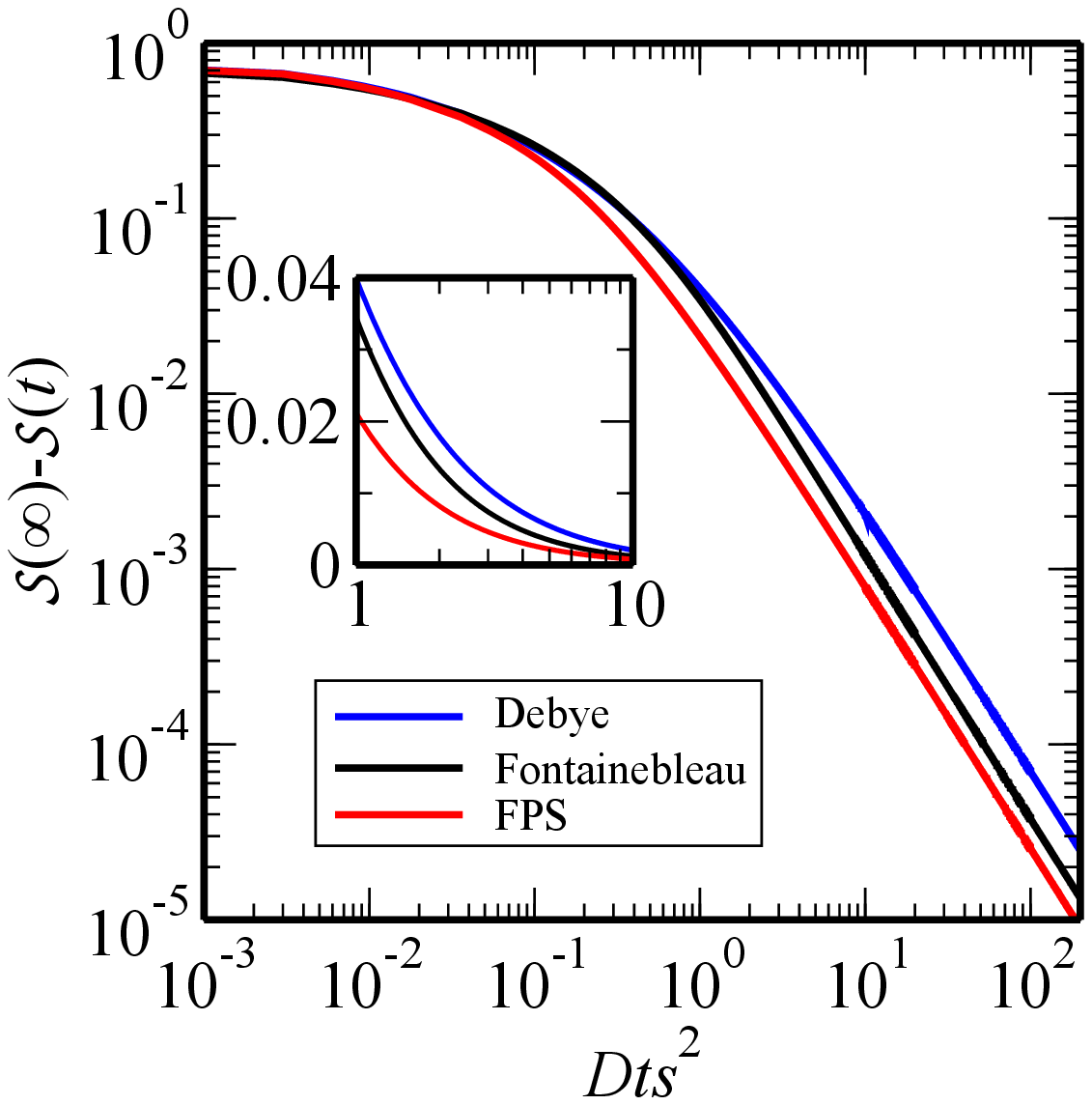}
  \caption{Spreadabilities for Fontainebleau sandstone, Debye random media and fully penetrable spheres with $\phi_2=0.157$ at small, intermediate and large dimensionless times $0.001<Dts^2<200$. The inset plots on a linear scale these spreadabilities at intermediate and large dimensionless times $1<Dts^2<10$.}
  \label{fontainebleau_spr}
\end{figure}

Figure \ref{fontainebleau_spr} shows the spreadabilities for Debye random media, Fontainebleau sandstone and FPS with $\phi_2=0.157$ at small, intermediate and large dimensionless times $0.001<Dts^2<200$, and the inset shows these spreadabilities at intermediate and large dimensionless times $1<Dts^2<10$. The spreadability for the sandstone closely resembles that for the Debye random media at small to intermediate dimensionless times ($0\leq Dts^2 \leq 1$), which reflects their similar small-scale correlations. We applied the algorithm described in Sec. \ref{algo} to extract large-time behaviors of these spreadabilities. As noted earlier, we treat the spreadability data for the sandstone as directly given when applying the algorithm. The extracted $\varphi$ value for the Fontainebleau sandstone is $\varphi=(3+\alpha)/2=1.501$, which indicates clearly that the sandstone is nonhyperuniform of the typical kind (Table \ref{set_in_3D}). The set-in time for the spreadability of the sandstone sample $Dt_S(10^{-6})s^2=24.9$ is also on the order of magnitude for typical nonhyperuniform media. Since all microstructures in Fig. \ref{fontainebleau_spr} are typical nonhyperuniform, they possess the same large-time power law decay behavior $\mathcal{S}(\infty)-\mathcal{S}(t)\sim Ct^{-3/2}$. However, their spreadabilities can be easily distinguished by their $C$ coefficients, which is directly proportional to $B=\tilde{\chi}_{_V}(0)$ of the corresponding spectral densities. As shown in Table \ref{set_in_3D}, the $C$ values extracted by our algorithm for Debye random media, Fontainebleau sandstone and FPS are 0.0686, 0.0379 and 0.0253, respectively. Since our algorithm typically gives less than 5\% error for extracted $C$ values, the aforementioned values of $C$ are significantly different. Moreover, the inset of Fig. \ref{fontainebleau_spr} clearly shows the distinction in the three spreadability curves due to their difference in $C$ coefficients. The trend of $C$ values for the three nonhyperuniform media confirms that the spectral density for the sandstone at small to intermediate wavenumbers lies in between those for Debye random media and FPS. 

\section{Conclusions}
\label{conc}
In this work, we investigated the practical applicability of the spreadability to probe microstructural information of real heterogeneous materials by accurately computing $\mathcal{S}(t)$ for idealized 2D and 3D models that represent a wide variety of hyperuniform and nonhyperuniform classes. These models include Debye random media, fully penetrable spheres, equilibrium hard spheres, which are nonhyperuniform, as well as sphere packings derived from ``perfect glasses'', uniformly randomized lattices, disordered stealthy hyperuniform point processes and Bravais lattices, which are hyperuniform. Our calculations confirm the previously predicted trends \cite{To21d} that the small-, intermediate and long-time behavior of $\mathcal{S}(t)$ sensitively captures the small-, intermediate- and long-range characteristics of the models. For typical nonhyperuniform models, the intermediate-time spreadability is slower for models with larger values of the coefficient $B=\tilde{\chi}_{_V}(0)$. Our study of the aforementioned models enabled us to devise an algorithm that efficiently and accurately extract large-scale structural characteristics from time-dependent diffusion behaviors, as measured by the spreadability, or equivalently, by NMR or MRI data. The algorithm also determines the ``set-in'' time of the large-time behavior of time-dependent diffusion measurements. Such analysis was used to characterize the large-scale behaviors of a sample Fontainebleau sandstone, which we show is typical nonhyperuniform. Therefore, we have demonstrated that the spreadability provides a simple and robust probe of crucial microstructural information across length scales of realistic heterogeneous media. Our proposed algorithm in Sec. \ref{algo} is shown to be capable of accurately extracting large-scale behaviors of real samples from time-dependent diffusion data.

Remarkably, in the case of URL packings, we found that $\mathcal{S}(\infty)-\mathcal{S}(t)$ has nearly exponential decay at small to intermediate times, but transitions to power-law decay at large $t$, i.e., $\mathcal{S}(t)$ clearly captures both the lattice-like order on small to intermediate length scales and the nonstealthy hyperuniform behavior on large length scales. We showed that the transition time $t^*$ that separates small- and large-time decay behaviors is well-defined and has a logarithmic divergence as $b\rightarrow 0$. The behaviors of the spreadability for URL packings stand in sharp contrast with those for disordered stealthy hyperuniform packings, which are characterized by a relatively slow small-time decay and an exponentially fast large-time decay.

We remark that the spreadability formula derived in Ref. \cite{To21d} can be regarded as a type of Gaussian smoothing of the spectral density, since it is weighted by a Gaussian function \cite{To02a}; see Eq. (\ref{spreadability_fourier}). Indeed, we observe that the numerically obtained spectral densities in this study contain some degree of noise at small wavenumbers [Fig. \ref{chi_spec}(b)]. However, their corresponding excess spreadabilities are very smooth functions of time and exhibit unambiguous power-law or exponential decay at large $t$ [Fig. \ref{3dspread}(a) and \ref{2dspread}(a)], which can be analyzed using our proposed algorithm to extract large-scale structural characteristics. Therefore, extracting the exponent $\alpha$ from spreadability offers a more robust and accurate means to compute it from real data compared to the direct numerical fitting of the small-$k$ data of the spectral density.

It is noteworthy that the microstructural characteristics extracted from time-dependent diffusion measurements are capable of predicting other effective properties that depend on two-point correlations. For example, Torquato recently derived a formula for the fluid permeability of porous media that depends on the spectral density ${\tilde \chi}_{_V}({\bf k})$, which has been shown to provide reasonably accurate permeability predictions of certain of hyperuniform and nonhyperuniform porous media \cite{To20a}. He used this formula to show that the dimensionless fluid permeability, i.e., (permeability multiplied by $s^2$), for stealthy hyperuniform media is expected to be lower relative to nonhyperuniform ones and that the BCC lattice packing minimizes the fluid permeability among common crystal lattices (simple cubic, FCC and BCC) at fixed porosity \cite{To20a}. Thus, the BCC lattice packing yields the fastest spreadability and relatively low values of the fluid permeability for packings of identical spheres at a given porosity.

Our findings on the spreadabilities for the idealized models in this study provide the basis for inverse design of materials with desirable time-dependent diffusion properties. Specifically, one could realize desired short-, intermediate- and large-time diffusion rates by constructing two-phase media with the corresponding values of $s$, $\alpha$ and $B$ (or $K$ in the case of stealthy media). Thus, a promising avenue for future study is the determination of microstructures of two-phase media that realize prescribed functional forms of $\mathcal{S}(t)$. This is an inverse problem that can be potentially tackled by adapting methods employed to construct microstructures with prescribed autocovariance functions and physical properties \cite{Ye98a,Za11f,Ma20b,sk21}.

\section*{Acknowledgement}
The authors thank Zheng Ma and Murray Skolnick for helpful discussions and the code for reconstructing Debye random media. Acknowledgement is made to the donors of the American Chemical Society Petroleum Research Fund under Grant No. 61199-ND9 and to the Air Force Office of Scientific Research Program on Mechanics of Multifunctional Materials and Microsystems under Grant No. FA9550-18-1-0514 for support of this research.

\appendix*
\renewcommand\thefigure{\Alph{figure}}   
\section{Characteristic functions for URL point processes}
\label{sec:urlfTilde}
\setcounter{figure}{0}    

\begin{figure}[!ht]
\centering
\subfloat[]{
    \centering\includegraphics[trim=0 0 0 -20, clip, width=8cm]{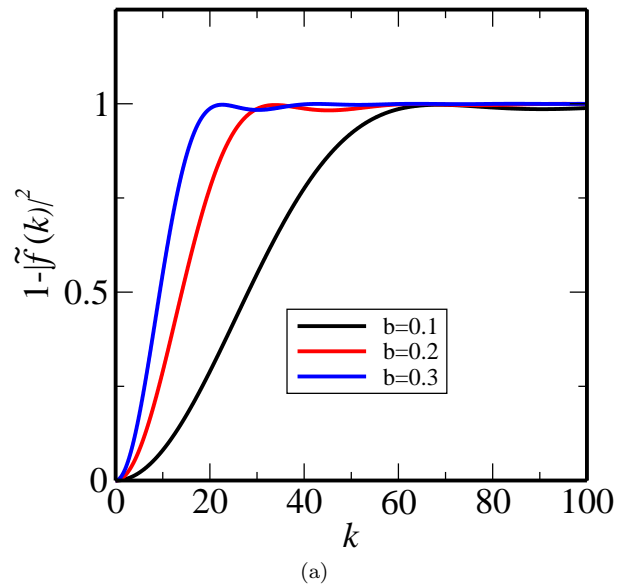}
}

\subfloat[]{
    \centering\includegraphics[trim=0 0 0 -20, clip, width=8cm]{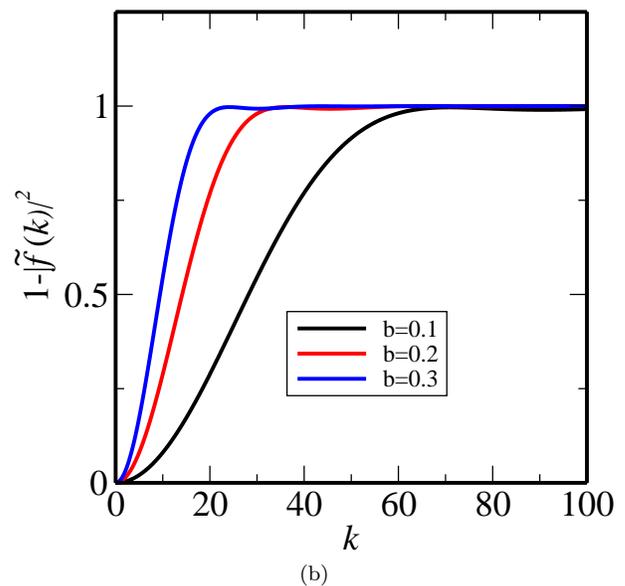}
}
  \caption{(a) Plot of $1-|\tilde{f}(k)|^2$ for 2D URL point processes. (b)  Plot of $1-|\tilde{f}(k)|^2$ for 3D URL point processes.}
  \label{fig:URL_fTilde}
\end{figure}

The characteristic function $\tilde{f}(\mathbf{k})$ of the uniform distribution on $[-b/2,b/2)^d$ is given by
\begin{equation}
    \tilde{f}(\mathbf{k})=\left(\frac{2}{b}\right)^d\prod_{i=1}^d\frac{\sin (bk_i/2)}{k_i},
    \label{eq:url_fTilde}
\end{equation}
where $k_i$ is the $i$-th component of the wavevector $\mathbf{k}$. Figure \ref{fig:URL_fTilde} shows plots of $1-|\tilde{f}(k)|^2$, i.e., the diffuse part of $S(k)$, for 2D and 3D URL point processes with $b=0.1,0.2$ and $0.3$. 
\color{black}
\newpage
\clearpage

%

\end{document}